\documentclass[a4paper,12pt]{article}

\usepackage{amsmath}
\usepackage{amsthm}
\usepackage{amssymb}
\usepackage{bm}
\usepackage{rotating}
\usepackage{color}
\usepackage{graphicx}
\usepackage{multirow}
\usepackage{booktabs,comment}
\usepackage[round, authoryear]{natbib}
\usepackage{pdfpages}
\usepackage[left = 2.2cm, right = 2.2cm, top = 2.7cm, bottom = 2.7cm]{geometry}

\newcommand*{\bb}{\boldsymbol}

\newcommand*{\hder}[2]{\partial{#1}/\partial{#2} }

\newcommand{\sign}[1]{\mathop{\rm sgn}(#1)}

\newtheoremstyle{example}
{3pt} 
{3pt} 
{} 
{0\parindent} 
{\bf}
{:} 
{.5em} 
{} 
\newtheoremstyle{theorem}
{3pt} 
{3pt} 
{\em} 
{0\parindent} 
{\bf}
{:} 
{.5em} 
{} 
\theoremstyle{example}
\newtheorem{example}{Example}[section]
\theoremstyle{theorem}

\title{Model-based clustering using copulas with applications} \date{\today}

\author{Ioannis Kosmidis \\ Department of Statistical Science,
  University College London \\ Gower Street, WC1E 6BT, London, UK  \\
  and \smallskip \\
  Dimitris Karlis \\
  Department of Statistics, Athens University of Economics and
  Business \\
  76 Patision Str, 10434, Athens, Greece}

\begin{document}

\def\var{\rm Var}

\maketitle

\begin{abstract}
  The majority of model-based clustering techniques is based on
  multivariate Normal models and their variants.  In this paper
  copulas are used for the construction of flexible families of models
  for clustering applications.  The use of copulas in model-based
  clustering offers two direct advantages over current methods: i) the
  appropriate choice of copulas provides the ability to obtain a range
  of exotic shapes for the clusters, and ii) the explicit choice of
  marginal distributions for the clusters allows the modelling of
  multivariate data of various modes (either discrete or continuous)
  in a natural way.  This paper introduces and studies the framework
  of copula-based finite mixture models for clustering
  applications. Estimation in the general case can be performed using
  standard EM, and, depending on the mode of the data, more efficient
  procedures are provided that can fully exploit the copula structure.
  The closure properties of the mixture models under marginalization
  are discussed, and for continuous, real-valued data parametric
  rotations in the sample space are introduced, with a parallel
  discussion on parameter identifiability depending on the choice of
  copulas for the components. The exposition of the methodology is
  accompanied and motivated by the analysis of real and artificial
  data.  
\end{abstract}

\section{Introduction}
\label{intro}
\subsection{Finite Mixture models for real-valued data}
The use of finite mixture models in clustering is finding a large
number of applications, mainly because it allows standard
statistical modelling tools to be used in order to assess and
evaluate the clustering. The density or probability mass function
of a finite mixture model is defined as
\begin{equation}
  \label{mixture_model}
  h({\bb x}; {\bb \theta}, {\bb \pi}) = \sum_{j = 1}^k \pi_j f_j({\bb
    x}; {\bb \theta}_j)  \quad ({\bb x} \in \Re^p) \, ,
\end{equation}
where ${\bb \theta} = ({\bb \theta}_1^\top, \ldots, {\bb
  \theta}_k^\top)^\top \in \Theta_1 \times \ldots \times \Theta_k$,
and $\pi_j \in (0,1)$ with $\sum_{j = 1}^k \pi_j = 1$.  Appropriate
choices of $f_j({\bb x}; {\bb \theta}_j)$ can result in flexible
models of small complexity.  \cite{BanfieldRaftery:93} and the book of
\cite{mclachlan:00} provide a detailed treatment of the framework of
finite mixture modelling for clustering.

For continuous data, a common choice for the component densities
$f_j({\bb x}; {\bb \theta}_j)$ $(j = 1, \ldots, k)$ is the density of
the multivariate Normal distribution. This is mainly because of the
convenience it offers in estimation (closed-form maximization steps in
the EM algorithm) and interpretation (easy marginalization for
visualising fitted components and the mixture density). The resultant
clusters, though, are limited to be elliptical in shape, and as is
demonstrated in
\citet{hennig:10}, one may need more than one multivariate Normal
components, in order to fit a single non-elliptical cluster.

\begin{figure}[t!]
  \begin{center}
    \includegraphics[width =
    0.99\textwidth]{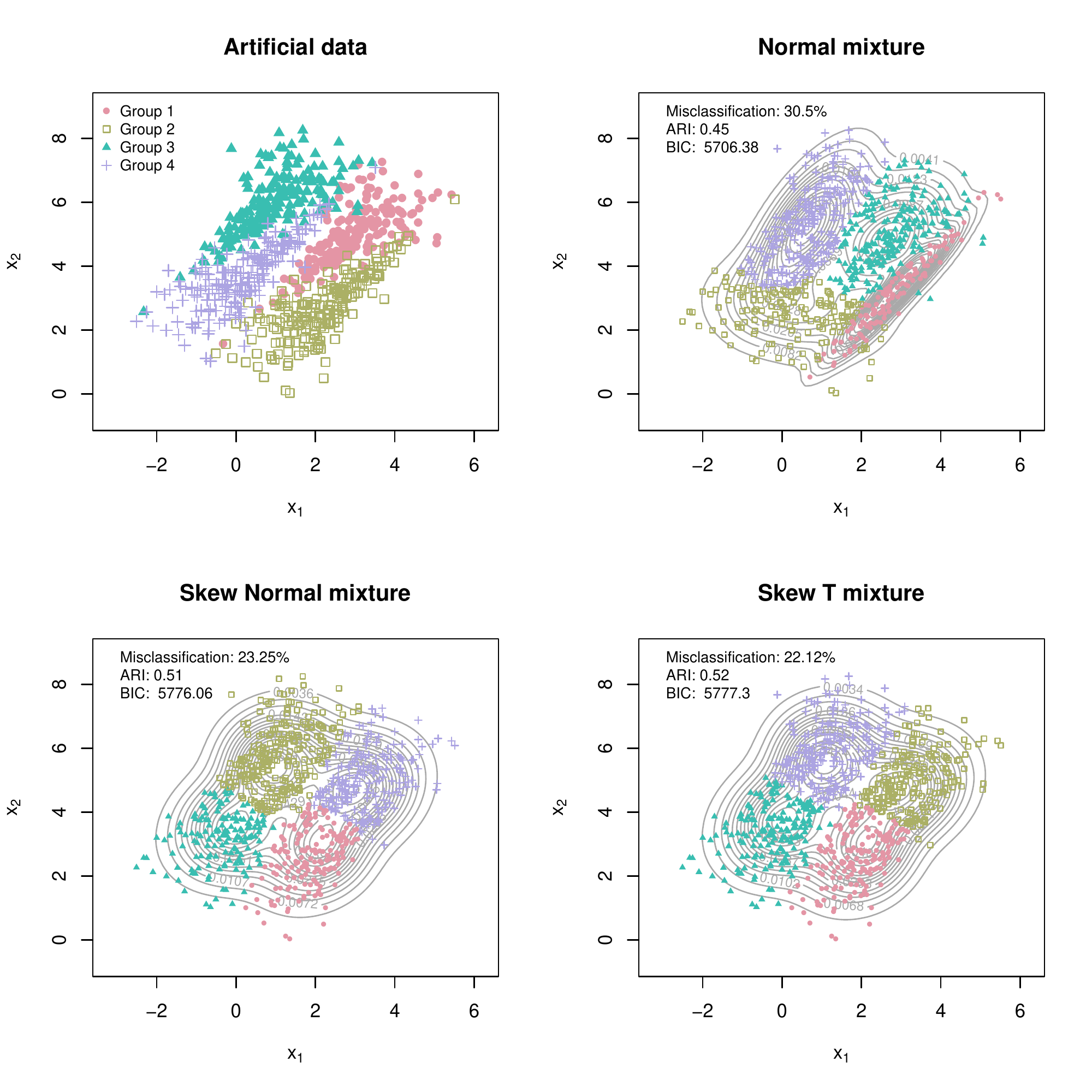}
  \end{center}
    \caption{An artificial data set with observations on two continuous
    variables (top left), a fitted mixture of four 2-dimensional
    Normal distributions (top right), a fitted mixture of four
    2-dimensional skew-Normal distributions (bottom left) and a fitted
    mixture of four 2-dimensional skew-t distributions (bottom
    right).}
  \label{motivatingFig}
\end{figure}

Such restrictions of multivariate Normal finite mixtures have
resulted in an expanding literature where other special component
distributions are considered.  Prominent examples of alternative
component densities include multivariate $t$ distributions
\citep[see,][]{andrews:11}, multivariate skew-Normal and skew-$t$
distribution \citep[see, for example,][]{ fruhwirth:10, lee:14},
multivatiate skew student-$t$-Normal distributions
\citep{lin:14}, multivariate Normal inverse Gaussian distributions
\citep{karlis:09}. Other attempts can be found in \citet{forbes:14}
for finite mixtures of multivariate scaled Normal distributions and
\citep{Morris20132088} for mixtures of shifted asymmetric Laplace
distributions. The results of such studies indicate that the
introduction of heavy tails and/or skewness allows the construction of
more parsimonious models than multivariate Normal mixtures, which can
also bridge the gap between the number of clusters present in the data
and the number of components used in the mixture.

Despite the added flexibility that such mixture models offer, all of
them force the data to obey very specific marginal properties, and
they are not appropriate, for example, in cases where the simultaneous
treatment of real-valued observations, strictly positive and
observations in $(0,1)$ is needed. In such cases one needs to either
ignore the range of the variables and treat them as real-valued or
apply appropriate transformations that map the original range of the
observations on the real line. Furthermore, even for real-valued
variables, as Example~\ref{motivatingExample} illustrates, current
methods can fail to capture certain dependence structures.

\begin{example}
  \label{motivatingExample} Consider the artificial data set shown in
  the top left plot of Figure~\ref{motivatingFig}. The data set is
  formed by four distinct clusters of observations each shown in a
  different colour on the plot. A Clayton and a survival Clayton
  copula with Normal marginals has been used for generating Groups 3
  and 4, and then an exact copy of the latter has been translated
  appropriately in order to form Groups 1 and 2.

  In an attempt to reconstruct the true groups, the data set was
  fitted using a bivariate Normal mixture model, a bivariate
  skew-Normal mixture model, and a bivariate skew-t mixture model.
  All fitting procedures were initialized by the best k-means
  clustering in four clusters after 1000 random starting points. The
  resulting classification plots are shown in
  Figure~\ref{motivatingFig}. Each plot also provides the value of the
  Bayesian Information Criterion (BIC) for each model and the
  corresponding misclassification rate. As is apparent none of the
  three models performs well in detecting the true shape of the
  underlying clusters with the misclassification rates ranging between
  $22.12\%$ to $30.5\%$ and adjusted Rand index (ARI) between $0.45$
  and $0.51$.

  The challenge with the artificial data set in
  Figure~\ref{motivatingFig} is the tail behaviour that the true
  groups demonstrate. If we restrict the number of components to four,
  the demonstrated extreme tail dependence and the small distances
  between the true groups makes models based on elliptical components
  (like Normal mixtures) incapable of capturing the true shape of
  the clusters. Moreover, in this example, models that are based on
  non-elliptical components (like skew-Normal and skew-t
  distributions) seem to be not flexible enough to capture the true
  characteristics of the data. \qed
\end{example}

\vspace{0.5cm}

\subsection{Finite mixture models for clustering non-continuous data}
For non-continuous data, one needs to specify $f_j({\bb x}; {\bb
  \theta}_j)$ $(j = 1, \ldots, k)$ in (\ref{mixture_model}) through
probability mass functions. While there is a wealth of choices for
univariate non-continuous distributions, the use of multivariate
non-continuous distributions for the definition of mixture models is
limited due to the difficulty in constructing easy to work with models
that allow practical flexibility on the dependence structure. Some
successful, but limited in application examples, are finite mixtures
of multivariate Poisson distributions \citep{karlis:07}, finite
mixtures of multinomial distributions \citep{jorgensen:04} and models
based on conditionally independent Poisson distributions \citep[see,
for example][]{Alfo2011185}. Mixture models with latent structures
have been considered in \citet{Browne20122976}, but these can have
limitations because of assumptions like conditional independence.

\subsection{Setting and fitting flexible finite mixture models}
Subsections 1.1 and 1.2 highlight the need for a new framework for
setting and fitting mixture models, which can i) match the flexibility
that current proposals offer, and ii) can accommodate the modelling of
data with either continuous or non-continuous domains.

Copulas offer the means for constructing such a framework; their
extensive use in the modelling of applications with multivariate
data is due to the flexibility they offer in describing dependence
and in that they allow the construction of multivariate models
with prescribed marginals. \citet{nelsen:06} provides an
introduction to the concept of copulas.
Moreover,
specifically for continuous data, common dependence measures like
Kendall's $\tau$ and Spearman's $\rho$ are marginal-free and
depend solely on the copula. This fact allows the easy
construction of multivariate mixture models for continuous data by
first selecting the marginal properties of the variables involved
and then the dependence structure implied by the mixture
components.

A few  attempts have already been made in the direction of
facilitating the flexibility that copulas offer in model-based
clustering \citep[see, for example][]{jajuga:06,
  lascio:12, vrac:12}. The current paper sets a thorough framework for
constructing mixture models using copulas, highlighting the
benefits but also the challenges of their use in practice. The
ingredients for constructing copula-based mixture models are
described in Section~2. Section~3 provides the details for maximum
likelihood estimation through Expectation-Maximization (EM)
algorithm  and proposes relevant procedures for obtaining starting
values from the combination of a partitioning algorithm (like
$k$-medoids) and of component-wise applications of the Inference
Functions from Margins (IFM) method of
\citet[][Chapter~10]{joe:97}. Section~4 focuses on the case of
modelling continuous multivariate data. The special structure of
the complete-data log-likelihood is exploited for producing more
efficient and stable variants of the Maximization step of the EM
algorithm. Topics like the modelling of mixed-domain continuous
data are discussed and a novel extension of the standard
copula-based mixture model is presented that applies parametric
component-wise rotations in the sample space and has effortless
implementation. Section~5 examines the property of closure under
marginalisation for copula-based mixture models and Section~6
provides a description of the use of the framework for modelling
multivariate discrete data. The challenges in estimation and model
specification compared to the continuous case are discussed.  The
exposition of the methodology is accompanied and motivated by the
analysis of real and artificial data. The paper concludes in
Section~7 with a discussion including descriptions of new research
directions that the current work offers.

\section{A flexible specification of mixture models}
\label{mixturemodels}

\subsection{Mixture models through copulas}
A copula $C(u_1, \ldots, u_p)$ is a distribution function with uniform
marginals. The importance of copulas in statistical modelling stems
from Sklar's theorem \citep[see,][\S 2.3]{nelsen:06}, which
shows that every multivariate distribution can be
represented via the choice of an appropriate copula and, more
importantly, it provides a general mechanism to construct new
multivariate models in a straightforward manner.

The copula-based mixture model is defined as in (\ref{mixture_model})
but now ${\bb \theta}_j$ is partitioned as $({\bb \gamma}_j^\top, {\bb
  \psi}_j^\top)^\top$ and $f_j({\bb x}; {\bb \theta}_j)$ is the density (or
probability mass function) corresponding to a distribution function
\begin{equation}
  \label{basic_model}
  F_j({\bb x}; {\bb \psi}_j, {\bb \gamma}_j) = C_j(G_1(x_1; {\bb \gamma}_{j1}), \ldots, G_p(x_p;
  {\bb \gamma}_{jp}); {\bb \psi}_j) \quad (j = 1, \ldots, k)\, ,
\end{equation}
where $G_1, \ldots, G_p$ are univariate marginal cumulative
distribution functions.  As far as the model parameters are concerned,
${\bb \gamma}_j$ contains the parameter vectors ${\bb \gamma}_{jt}$
for all marginals for $j$th component $(t = 1, \ldots, p)$ and ${\bb
  \psi}_j$ contains the parameters of the copula used for the $j$-th
component.


\subsection{Construction of mixture models for any type of marginals}
\label{differenttypes}

The definition of the component density $F_j$ through the choice
of a copula $C_j$ and the choice of marginal distributions $G_1,
\ldots, G_p$ leads to a flexible framework for model-based
clustering that according to Sklar's theorem necessarily
encompasses all known mixture models and allows the convenient
construction of new mixture models that can handle any of
continuous, discrete  data.

Temporarily omitting the component index and suppressing the
dependence on the parameters, assume that the density of the copula
$C(u_1, \ldots, u_p)$ exists and is $c(u_1, \ldots, u_p)
= \partial^pC(u_1, \ldots, u_p)/\partial u_1\ldots \partial u_p$.
Then the component density for continuous marginals is
\[
f({\bb x}) = c(G_1(x_1), \ldots, G_p(x_p))\prod_{t = 1}^p g_t(x_t) \, ,
\]
where $g_t(x) = dG_t(x)/dx$ is the density function for the $t$th
marginal distribution. For discrete data, the probability mass
function is given in \citet[expression~(1.2)]{panagiotelis:12}, and
results from finite differences of the distribution function as
\begin{equation}
  \label{copDiscrete}
  P({\bb x}) = \sum_d \sign{{\bb d}} C(G_1(d_1), \ldots, G_p(d_p))
  \, ,
\end{equation}
with ${\bb d} = (d_1, \ldots, d_p)$ vertices, where each $d_t$ is
equal to either $x_t$ or $x_t - 1$ $(t = 1, \ldots, p)$, and
\[\sign{{\bb d}} = \left\{
  \begin{array}{rc} 1\,, & \text{if } d_t = x_t - 1 \text{ for an even
      number of } t\text{'s} \\ -1\,, & \text{if } d_t = x_t - 1
    \text{ for an odd number of } t\text{'s} \\
  \end{array} \right.\, .
\]

The model defined from (\ref{mixture_model}) and (\ref{basic_model})
being a finite mixture allows for inferential procedures based on the
standard theory of finite mixtures, like use of the
EM algorithm for maximum likelihood estimation
and the use of model selection criteria.

\section{Model fitting}
\label{modelfitting}

\subsection{Full Expectation Maximization algorithm}
\label{EMsection}
Suppose that a sample of $n$ $p$-vectors ${\bb x}_1, \ldots, {\bb
  x}_n$ is available, which are assumed to be realizations of
independent random variables ${\bb X}_1, \ldots, {\bb X}_n$ each
with distribution with density or probability mass function as
defined by (\ref{mixture_model}) and (\ref{basic_model}). The
maximization of the likelihood function based on that sample can
be performed using the EM algorithm. At the $\ell$th iteration of
the algorithm: 
\begin{itemize}
\item {\em E-step:} Calculate
  \[ w_{ij}^{(\ell + 1)} = \frac{\pi_j^{(\ell)} f_j({\bb x}_i; {\bb
      \theta}_j^{(\ell)})}{ \sum_{j = 1}^k \pi_j^{(\ell)} f_j({\bb
      x}_i; {\bb\theta_j}^{(\ell)})} \quad (i = 1, \ldots, n;~~ j = 1,
  \ldots, k) \, .
  \]
\item {\em M-step 1:} Set $\pi_j^{(\ell + 1)} = \sum_{i = 1}^n
  w_{ij}^{(\ell + 1)}/n$ $(j = 1, \ldots, k)$.
\item {\em M-step 2:} Maximize
  \[
  \sum_{j = 1}^k
  \sum_{i = 1}^n w_{ij}^{(\ell + 1)} \log\left\{f_j({\bb x}_i; {\bb
      \theta}_j) \right\} \, ,
  \] with respect to ${\bb \theta}$ to obtain an updated value ${\bb
    \theta}^{(\ell + 1)}$ for the copula and marginal parameters.
\end{itemize}

The algorithm iterates between the {\em E-step} and the {\em M-step}
until some convergence criterion is satisfied. In all the examples in
the current paper the termination criterion that is used is that the
relative increase
$\{l(\theta^{(\ell+1)}, \pi^{(\ell+1)}) - l(\theta^{(\ell)},
\pi^{(\ell)})\}/l(\theta^{(\ell)}, \pi^{(\ell)})$
of the log-likelihood $l(\theta, \pi)$ in two successive iterations is
less than $\epsilon = 10^{-8}$.

\subsection{Computational details}
\label{compute}
\subsubsection{Maximization step}
For the general model defined by (\ref{mixture_model}) and
(\ref{basic_model}), {\em M-step 2} of the EM iteration is
generally not available in closed-form and needs to be performed
numerically. At the current level of generality, it is recommended
to take advantage of the separable form of the complete-data
log-likelihood for mixture models, which allows to break down the
maximization task into $k$ independent maximizations of weighted
likelihoods
\[ {\bb\theta}_j^{(\ell+1)} = \arg\max_{\Theta_j} \sum_{i = 1}^n
w_{ij}^{(\ell + 1)} \log\left\{f_j({\bb x}_i; {\bb\theta}_j)
\right\}\quad (j = 1, \ldots, k) \, ,
\]
that can be performed in parallel.

\subsubsection{Starting values}
\label{starting}
For calculating the starting values for ${\bb \pi}$ and ${\bb\theta}$
the following procedure is proposed which takes into account both the
copula and the marginal specification of each component in the mixture
model. The procedure is an application of the Inference Functions from
Margins (IFM) method \citep[][Chapter~10]{joe:97} for each component,
and relies on an initial classification vector that partitions the
observation indices $A = \{1, \ldots, n\}$ into exclusive subsets
$S_1, \ldots, S_k$, with $\cup_{j = 1}^kS_j = A$, of cardinality
$N_1, \ldots, N_k$, respectively. More specifically, the procedure for
obtaining starting values consists of the following steps:
\begin{enumerate}
\item[S1] Set the starting values for ${\bb \pi_j}$ using
${\pi^*_j} =
  N_j/n$ $(j = 1, \ldots, k)$.
\item[S2] Use maximum likelihood to fit the marginal $g_t$ on data
  $x_{it}$ for $i \in S_j$ in order to obtain starting values ${\bb
    \gamma}_{jt}^*$ for ${\bb \gamma}_{jt}$ $(t = 1, \ldots, p)$.
\item[S3] Use maximum likelihood to fit the copula $C_j(u_1, \ldots,
  u_p; {\bb \psi}_j)$ on observations $u_{it} = G_t(x_{it}; {\bb
    \gamma}_{jt}^*)$ $(i \in S_j;~~t = 1, \ldots, p)$, in order to get
  starting values ${\bb \psi}_j^*$ for the copula parameters ${\bb
    \psi}_j$.
\end{enumerate}

The initial classification vector can be obtained either using a
hard-partitioning distance-based algorithm (like $k$-means for
continuous data or $k$-medoids more generally) or by randomly sampling
$k$ observations and using the minimum distance of each to all other
observations in order to form $S_1, \ldots, S_k$.

\subsubsection{Choice of component ordering}
The possibility of using different copulas for the components of the
mixture model defined by (\ref{mixture_model}) and
(\ref{basic_model}), and the fact that the likelihood function for
mixture models generally has local maxima, make the solution of the EM
algorithm to depend on the order that the copulas appear in the
mixture.

A solution to this problem is to fit all models that result from all
possible permutations of the component copulas. Then, for each one of
the unique permutations of the components, the procedure in
Subsection~\ref{starting} is applied, taking care to use the same
initial classification vector (and labelling) for $A$ across
permutations. Then the fitted model with the largest value for the
maximized log-likelihood is
chosen. Example~\ref{motivatingExampleCont} below, uses this procedure
for the choice of component ordering. Subsection~\ref{goodfits}
presents an alternative, less intensive solution, which can give rise
to flexible mixture models without the need of considering many
different candidate copulas for the components. That solution is based
on extending the specification of the mixture model by allowing for
component-wise parametric rotations.

If the same copula is used across the components of the mixture model,
then the sensitivity that the result of the EM algorithm can have on
the starting values can be alleviated by trying several of those. This
can be done by choosing a number of sets of $k$ randomly sampled
observations, and construct distinct classification vectors by minimum
distance, as described in Subsection~\ref{starting}. Then, for each
vector, component-wise IFM is used to get the corresponding set of
starting values and initialize the EM iterations. The model fit with
the largest maximized log-likelihood is the one that is kept. The
above process is used to carry out the analyses in Example~\ref{nba}
and Example~\ref{cognitive}.

\section{Continuous data}

\subsection{Maximization step}
\label{maxContinuous}
For the analysis of continuous data, {\em M-step 2} in
Subsection~\ref{EMsection} takes the form
\begin{itemize}
\item {\em M-step 2:} Maximize the log-likelihood
  \begin{equation}
    \label{M2continuous}
    \sum_{j = 1}^k \sum_{i = 1}^n
    w_{ij}^{(\ell + 1)} \left[ \log c_j(G_1(x_{i1}; {\bb \gamma}_{j1}), \ldots,
      G_p(x_{ip}; {\bb \gamma}_{jp}); {\bb \psi}_j) +
      \sum_{t = 1}^p\log g_t(x_{it}; {\bb \gamma}_{jt})  \right] \, ,
  \end{equation}
\end{itemize}
with respect to ${\bb \psi}_1, \ldots, {\bb \psi}_k, {\bb
  \gamma}_{11}, \ldots, {\bb \gamma}_{1p}, {\bb \gamma}_{k1}, \ldots,
{\bb \gamma}_{kp}$, where ${\bb \gamma}_{jt}$ is the vector of
parameters of the $t$th marginal distribution for the $j$th component
of the mixture $(t = 1, \ldots, p; j = 1, \ldots, k)$.

As is apparent from (\ref{M2continuous}) the only necessary
ingredients for implementing the EM algorithm for mixtures of copulas
for continuous data are the specification of the copula densities
$c_1, \ldots, c_k$ and the specification of the marginal density and
distribution functions $g_1, \ldots, g_p$ and $G_1, \ldots, G_p$ ,
respectively.


The particular form of the complete data log-likelihood for continuous
data allows here the use of the Expectation/Conditional Maximization
(ECM) algorithm of \citet{meng:93}, where the full maximization of the
complete data log-likelihood is relaxed to maximization in blocks;
first with respect to the marginal parameters given the current value
of the copula parameter and then with respect to the copula parameter
given the updated values for the marginal parameters. In mathematical
notation, {\em M-step 2} in Subsection~\ref{EMsection} is replaced by
the steps
\begin{itemize}
\item {\em CM-step 1:} Maximize
  \begin{equation}
    \label{CM1continuous}
    \sum_{j = 1}^k \sum_{i = 1}^n
    w_{ij}^{(\ell + 1)} \left[ \log c_j(G_1(x_{i1}; {\bb \gamma}_{j1}), \ldots,
      G_p(x_{ip}; {\bb \gamma}_{jp}); {\bb \psi}^{(\ell)}_j) +
      \sum_{t = 1}^p\log g_t(x_{it}; {\bb \gamma}_{jt})  \right] \, ,
  \end{equation}
  with respect to ${\bb \gamma}_{11}$, $\ldots$, ${\bb \gamma}_{1p}$,
  ${\bb \gamma}_{k1}$, $\ldots$, ${\bb \gamma}_{kp}$ to obtain updated
  values ${\bb \gamma}^{(\ell + 1)}_{11}$, $\ldots$, ${\bb
    \gamma}^{(\ell + 1)}_{1p}$, ${\bb \gamma}^{(\ell + 1)}_{k1}$,
  $\ldots$, ${\bb \gamma}^{(\ell + 1)}_{kp}$ for the marginal
  parameters.
\item {\em CM-step 2:} Maximize
  \begin{equation}
    \label{CM2continuous}
    \sum_{j = 1}^k \sum_{i = 1}^n
    w_{ij}^{(\ell + 1)} \left[ \log c_j(G_1(x_{i1}; {\bb \gamma}^{(\ell + 1)}_{j1}), \ldots,
      G_p(x_{ip}; {\bb \gamma}^{(\ell + 1)}_{jp}); {\bb \psi}_j)\right] \, ,
  \end{equation}
  with respect to ${\bb \psi}_1, \ldots, {\bb \psi}_k$ to obtain updated
  values ${\bb \psi}^{(\ell + 1)}_1, \ldots, {\bb \psi}^{(\ell +
    1)}_k$ for the copula parameters.
\end{itemize}
According to the definitions and results in \citet{meng:93}, the ECM
algorithm that results by replacing {\em M-step 2} with the pair {\em
  CM-step 1} and {\em CM-step 2} shares all the convergence properties
of the full EM algorithm, and, in this particular case, is
more computationally efficient and stable, because {\em CM-step 2} consists
of a simple maximization with respect to the copula parameters.
Furthermore, {\em CM-step 1} and {\em CM-step 2} can each
be broken down into parallel optimizations across components, as in
the case of the full EM in Subsection~\ref{compute}, which
significantly reduces computation time in multicore systems.

The pair of {\em CM-step 1} and {\em CM-step 2} is similar to the IFM
method for fitting the complete data likelihood. Their difference lies
in {\em CM-step 1} where instead of maximizing the weighted sum of
marginal log-likelihoods, a valid ECM algorithm requires the
maximization of a penalized version of it where the penalty depends on
the log copula density at the current value for the copula parameter.

\begin{example}
  \label{motivatingExampleCont} Consider the setting of
  Example~\ref{motivatingExample}. The Normal mixture model failed
  to capture the dependence structure that is apparent in the true
  groups because of the strict elliptical shape of the component
  densities. Furthermore, two of the fashionable methods that allow
  non-elliptical clusters (mixtures of skew-Normals and skew-t
  distributions) were not able to recover the dependence structure of
  the groups in the artificial data.

  The Gumbel copula and the Clayton copula can capture varying degrees
  of upper and lower tail dependence respectively and for the purposes
  of this example we consider a mixture model of two Gumbel copulas
  and two Clayton copulas with Normal marginals. The Gumbel copula is
  defined as
  \begin{equation}
    \label{gumbelCopula}
    C^{(G)}(u_1, u_2 ; \psi) = \exp\left[ - \left\{ (-\log u_1)^\psi + (-\log
        u_2)^\psi \right\}^{1/\psi} \right]\,, \quad \psi \in [1,
    \infty)\, ,
  \end{equation}
  and the Clayton copula is defined as
  \begin{equation}
    \label{claytonCopula}
    C^{(C)}(u_1, u_2; \psi) = \left(u_1^{-\psi} + u_2^{-\psi} - 1
    \right)^{-1/\psi}\,, \quad \psi \in (0, \infty) \, .
  \end{equation}
  The associated densities $c^{(G)}(u_1, u_2 ; \psi)$ and
  $c^{(C)}(u_1, u_2; \psi)$ can be obtained by direct differentiation
  of $C^{(G)}(u_1, u_2; \psi)$ and $C^{(C)}(u_1, u_2; \psi)$,
  respectively. The closed form expressions of those copula densities
  are given in \citet[][Corollary 1]{hofert:12} along with the
  corresponding expressions for other Archimedean copulas of arbitrary
  dimension. Then the density of the bivariate mixture model with two
  Gumbel and two Clayton components and Normal marginal distributions
  can be written as
  \begin{align}
    \label{artificialMixture}
    h({\bb x}; {\bb \theta}, {\bb \pi}) & = \sum_{j = 1}^2 \pi_j
    c^{(G)}\left\{\Phi\left(\frac{x_1 - \mu_{j1}}{\sigma_{j1}}\right),
      \Phi\left(\frac{x_2 - \mu_{j2}}{\sigma_{j2}}\right);
      \psi_j\right\} \prod_{t = 1}^2 \frac{1}{\sigma_{jt}}
    \phi\left(\frac{x_t - \mu_{jt}}{\sigma_{jt}}\right) \\ \notag & +
    \sum_{j = 3}^4 \pi_j c^{(C)}\left\{\Phi\left(\frac{x_1 -
          \mu_{j1}}{\sigma_{j1}}\right), \Phi\left(\frac{x_2 -
          \mu_{j2}}{\sigma_{j2}}\right); \psi_j\right\} \prod_{t =
      1}^2 \frac{1}{\sigma_{jt}} \phi\left(\frac{x_t -
        \mu_{jt}}{\sigma_{jt}}\right) \, ,
  \end{align}
  where $~~{\bb \theta} =~~ (\mu_{11}, \sigma_{11}, \mu_{12}, \sigma_{12},
  \psi_1, \ldots, ~~\mu_{41}, \sigma_{41}, \mu_{42}, \sigma_{42},
  \psi_4)^\top$ and $~~{\bb \pi} = (\pi_1, \ldots, \pi_4)^\top~~$ with \newline  ${\sum_{j
    = 1}^4 \pi_j = 1}$. The functions $\Phi(.)$, $\phi(.)$ are the
  distribution and density function of a standard Normal random
  variable, respectively.

  Then {\em M-step 2} of the maximization step at the $\ell$th
  iteration of the EM algorithm in Subsection~\ref{EMsection} consists
  of the maximization of each one of
  \begin{equation}
  \sum_{i = 1}^n w_{ir}^{(\ell + 1)} \left[ \log
    c\left\{\Phi\left(\frac{x_{i1} - \mu_{r1}}{\sigma_{r1}}\right),
      \Phi\left(\frac{x_{i2} - \mu_{r2}}{\sigma_{r2}}\right);
      \psi_r\right\} - \sum_{t = 1}^2\log\sigma_{rt} +
    \frac{1}{2}\sum_{t = 1}^2 \left(\frac{x_{it} -
        \mu_{rt}}{\sigma_{rt}}\right)^2 \right]
        \label{maxfun}
  \end{equation}
  for $r \in \{1, 2, 3, 4\}$, where $c \equiv c^{(G)}$ for $r\in \{1,
  2\}$ and $c \equiv c^{(C)}$ for $r\in \{3, 4\}$. For deriving the ECM
  algorithm in Subsection~\ref{maxContinuous}, each of those
  maximizations should be replaced by the maximization of the
  function given in (\ref{maxfun}), firstly on with respect to $\mu_{r1}, \sigma_{r1}, \mu_{r2},
  \sigma_{r2}$ at the current value $\psi^{(\ell)}_{r}$ of the copula
  parameter, in order to obtain updated marginal parameters $\mu^{(\ell +
    1)}_{r1}, \sigma^{(\ell + 1)}_{r1}, \mu^{(\ell + 1)}_{r2},
  \sigma^{(\ell + 1)}_{r2}$, and, then, the maximization of
  \[
  \sum_{i = 1}^n w_{ir}^{(\ell + 1)} \log
  c\left\{\Phi\left(\frac{x_{i1} - \mu^{(\ell +
          1)}_{r1}}{\sigma^{(\ell + 1)}_{r1}}\right),
    \Phi\left(\frac{x_{i2} - \mu^{(\ell + 1)}_{r2}}{\sigma^{(\ell +
          1)}_{r2}}\right); \psi_r\right\}\, ,
  \]
  with respect to $\psi_r$ to obtain an updated value $\psi^{(\ell +
    1)}_r$ for the copula parameter. The latter is simply a
  maximization with respect to the scalar parameter $\psi_r$ and can
  be performed using line search in the domain of definition of the
  copula parameter.

  \begin{table}[t!]
    \begin{center}
      \begin{tabular}{ccccc}
        \toprule
        & Component 1 &
        Component 2 &
        Component 3 &
        Component 4 \\ \midrule
        & Gumbel &
        Gumbel &
        Clayton &
        Clayton \\ \midrule
        Mixing probabilities & $\hat{\pi}_{1} = 0.24$ &
        $\hat{\pi}_{2} = 0.24$ &
        $\hat{\pi}_{3} = 0.25$ &
        $\hat{\pi}_{4} = 0.27$ \\ \cmidrule{2-5}
        \multirow{4}{*}{Marginal parameters} & $\hat{\mu}_{11} = 2.31$ &
        $\hat{\mu}_{21} = 0.35$ &
        $\hat{\mu}_{31} = 2.79$ &
        $\hat{\mu}_{41} = 0.78$ \\
        & $\hat{\sigma}_{11} = 0.95$ &
        $\hat{\sigma}_{21} = 0.97$ &
        $\hat{\sigma}_{31} = 1.00$ &
        $\hat{\sigma}_{41} = 1.02$ \\
        & $\hat{\mu}_{12} = 2.73$ &
        $\hat{\mu}_{22} = 3.76$ &
        $\hat{\mu}_{32} = 4.77$ &
        $\hat{\mu}_{42} = 5.77$ \\
        & $\hat{\sigma}_{12} = 1.03$ &
        $\hat{\sigma}_{22} = 1.05$ &
        $\hat{\sigma}_{32} = 1.05$ &
        $\hat{\sigma}_{42} = 1.07$ \\ \cmidrule{2-5}
        Copula parameters & $\hat{\psi}_{1} = 2.86$ &
        $\hat{\psi}_{2} = 2.85$ &
        $\hat{\psi}_{3} = 3.56$ &
        $\hat{\psi}_{4} = 3.24$ \\ \bottomrule
      \end{tabular}
    \end{center}
    \caption{Maximum likelihood estimates for the  parameters for the mixture~(\ref{artificialMixture}).}
    \label{artificialEstimates}
  \end{table}

  In the current example, the possible permutations of the copulas for
  the components are $\{G, G, C, C\}$, $\{G, C, G, C\}$, $\{G, C, C,
  G\}$, $\{C, G, G, C\}$, $\{C, G, C, G\}$, and $\{C, C, G, G\}$,
  where $G$ and $C$ stand for Gumbel and Clayton, respectively.
  Table~\ref{artificialEstimates} shows the estimates for the
  parameters for that permutation of copulas which resulted in the
  largest maximized log-likelihood.

  The resulting classification plot is shown in
  Figure~\ref{motivatingCopulaFig}. As is apparent the copula-based
  mixture model is performing very well in capturing the shape of the
  original clusters; the misclassification rate is $9.5\%$ and the BIC
  value ($5598.74$) has greatly improved from the models in
  Figure~\ref{motivatingFig}. The resultant clustering has ARI of 0.77
  which dominates the clusterings obtained in \ref{motivatingExample}.

  \begin{figure}[t!]
    \begin{center}
      \includegraphics[width =
      0.99\textwidth]{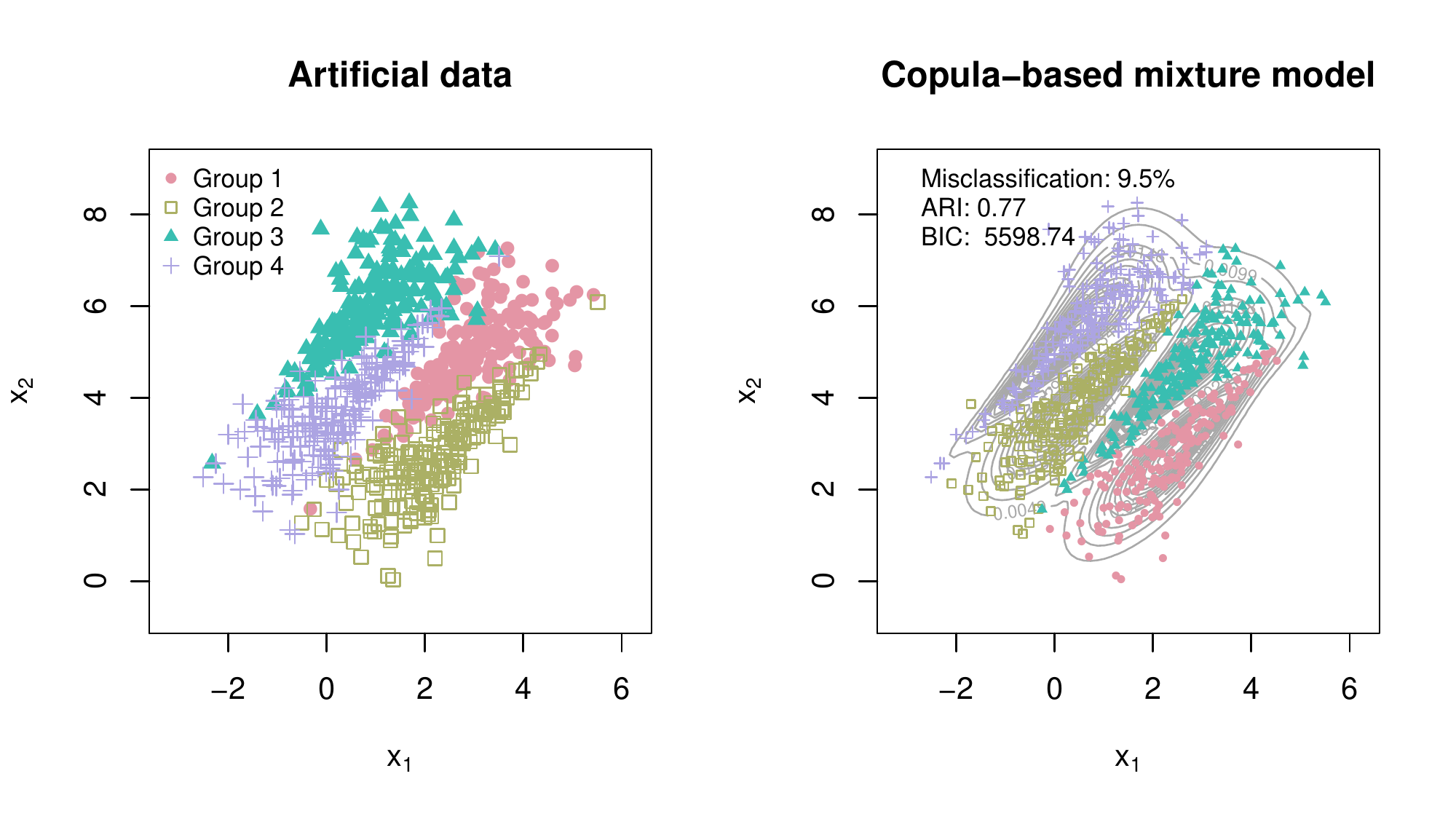}
    \end{center}
    \caption{The artificial data set of
      Example~\ref{motivatingExample} (left) and the contours of the
      fitted mixture of the bivariate mixture model with two Gumbel
      and two Clayton components and Normal marginal distributions
      (right).}
    \label{motivatingCopulaFig}
  \end{figure}

  \qed

\end{example}

\subsection{Bounded- and mixed-domain variables}
The decoupling of the dependence properties from the marginal ones
allows the easy construction of multivariate mixture models for
bounded-domain data (like percentages or strictly positive
variables).

Mixture models for such data are usually formed from component
densities defined by the product of independent univariate densities
of appropriate support. Those models imply that the univariate
marginal distributions of the component densities are independent
conditional on the component membership. In effect, such models
attempt to capture the dependence in the data only through the mixing
probabilities. For example, \citet{dean:13} define multivariate Beta
mixture models in this way, and use them highlighting that the implied
conditional independence assumption can be rather restrictive in
practice.

Under the current framework, models with more complex dependence can
be defined by simply setting the component copula densities $c_j$ in
(\ref{M2continuous}), and choosing marginals with the required
support.

More importantly, the current methodology can be used to construct
multivariate mixture models for mixed-domain
marginals. Example~\ref{nba} below considers a mixture model with
components that have seven Beta and one Gamma marginal distributions
each.

\begin{example}
  \label{nba}
  Hoopdata.com is a website that was launched in 2009 and provides an
  extensive database for NBA statistics.
  We used Hoopdata.com's database to gather data for the
  scoring behaviour of the $493$ NBA players that had more than $24$
  minutes game time (which accounts for half a game played in full) in
  the 2011-2012 season. For each player, the data has observations for
  the season free throws percentage (``FTper''), the field goals
  percentage (``FGper''), the three point percentage (``ThreePper''),
  the percentage of field goals assisted, the percentage blocked, the
  percentage of ``And1'' field goals, and the total points scored in
  hundreds (``PointsHun''). The aim of the analysis is to form groups
  of players in terms of their performance.

  For all players that had observation $0\%$ or $100\%$ in any of the
  percentages, the observation is replaced by $0.01\%$ and $99.99\%$,
  respectively. After making the substitutions, the free
  throws, the field goals and the three point percentages take values
  in $(0, 1)$. Furthermore, the points scored are all positive. Plausible
  marginal specifications are a Beta distribution for modelling each
  of the percentage variables and a Gamma distribution for modelling
  the points scored. Then we use a 7-variate Gaussian copula for
  modelling the dependence between the 7 variables.

  The multivariate Gaussian copula with correlation matrix $R$ is
  defined as
  \begin{equation}
    \label{multivariateNormal}
    C^{(N)}(u_1, \ldots, u_p; R) = \Phi_p(\Psi(u_1), \ldots, \Psi(u_p); R) \, ,
  \end{equation}
  where $\Phi_p(., \ldots, .)$ is the distribution function of a
  standard $p$-variate Normal distribution with correlation matrix $R$
  and $\Psi(.) = \Phi_1^{-1}(.)$ is the inverse distribution function
  of a standard Normal distribution. The matrix $R$ has the general
  form
  \begin{equation}
    \label{cormat}
    R = \left[
      \begin{array}{cccc}
        1 & \rho_{12} & \ldots & \rho_{1p} \\
        \rho_{12} & 1 & \ldots & \rho_{2p} \\
        \vdots & \vdots & \ddots & \vdots \\
        \rho_{1p} & \rho_{2p} & \ldots & 1 \\
      \end{array}
    \right] \, ,
  \end{equation}
  where $\rho_{tt'} \in [-1, 1]$ is the correlation between the $t$th
  and $t'$th variable $(t,t' \in \{1, \ldots, p\}; t \ne t')$.

  For the current case the density of the mixture model to be fitted
  is
  \begin{equation}
    \label{mixdensityBeta}
    \sum_{j = 1}^k \pi_j \phi_7\left[ \Psi\left\{G_{j1}(x_1)\right\},
      \ldots,
      \Psi\left\{G_{j7}(x_7)\right\}; R_j \right] \prod_{t = 1}^7
    \frac{g_{jt}(x_t)}{\phi_1[\Psi\{G_{jt}(x_t)\}]} \, ,
  \end{equation}
  where $\phi_p$ is the density of a $p$-dimensional standard Normal
  distribution and $g_{jt}(x) = \hder{G_{jt}(x)}{x}$, where for $t \in
  \{1, \ldots, 6\}$, $G_{jt}(x)$ is the distribution function of a Beta
  random variable with shape parameters $\alpha_{jt}$, and $G_{j7}(x)$
  is the distribution function of a Gamma random variable with shape
  parameter $\kappa_j$ and scale $1/\lambda_{j}$. More specifically,
  the density functions for the marginals are
  \[
  g_{jt}(z) = \left\{
    \begin{array}{ll}
      \displaystyle \frac{z^{\alpha_{jt} - 1} (1 -z)^{\beta_{jt} -
          1}}{B(\alpha_{jt}, \beta_{jt})} \, ,
      & t \in \{1, \ldots, 6\} \\ \\
      \displaystyle
      \frac{\lambda_{j}^{\kappa_{j}}}{\Gamma(\kappa_{j})} z^{\kappa_{j}
        - 1} \exp(- \lambda_{j} z) \, ,   & t = 7
    \end{array}
  \right.  \quad (j = 1, \ldots, k)\, .
  \]
  The matrix $R_j$ has exactly the same structure as the matrix $R$ in
  (\ref{cormat}) but the correlations depend on $j$ ensuring that each
  component in the mixture can accommodate different correlation
  structures. Hence, the parameters to be estimated are $\alpha_{j1}$,
  $\beta_{j1}$, $\ldots$, $\alpha_{j6}$, $\beta_{j6}$, $\kappa_j$ and
  $\lambda_j$ for the marginals of the $j$th component, $\rho_{12,
    j}$, $\ldots$, $\rho_{17, j}$, $\ldots$, $\rho_{67, j}$ for the
  copula of the $j$th component $(j = 1, \ldots, k)$ and the mixing
  proportions $\pi_1, \ldots, \pi_{k-1}$. Hence, the model in
  (\ref{mixdensityBeta}) has $q = 36k- 1$ free parameters. The number
  of free parameters can be further reduced by fitting models nested
  to (\ref{mixdensityBeta}) that have a structured correlation
  matrix. For example, a nested model to (\ref{mixdensityBeta}) with
  $16k - 1$ parameters can be formed by considering exchangeable
  correlation for each component where all correlations appearing in
  $R_j$ are equal to $\rho_j$ $(j = 1, \ldots, k)$.

  For $k = 2, \ldots, 9$, density (\ref{mixdensityBeta}) is fitted to
  the data on the NBA players with both unstructured and exchangeable
  correlation. For the models with exchangeable correlation and for
  each $k$, $15$ sets of starting values were obtained by selecting
  $15$ sets of $k$ randomly selected observations for the
  intialization of the component-wise IFM procedure in
  Subsection~\ref{starting}. Then the model with the highest
  log-likelihood was chosen and was also used to initialise the model
  with unstructured correlation.

  Furthermore, we restricted the variance of each of the Beta
  marginals involved in the mixture to be greater than $10^{-4}$. In
  this way, unbounded likelihood values relating to observations with
  the same percentage value can be avoided.

  Table~\ref{mixBetaResults} lists the maximized log-likelihood, the
  number of parameters $q$ and the BIC value for the best $18$ fitted
  mixtures. The model with exchangeable correlation matrix and $k = 6$
  has the lowest BIC value.

  \begin{table}[t!]
    \begin{center}
      \begin{small}
        \begin{tabular}{cccclcccl}
          \toprule
          k & Log-likelihood & $q$ & BIC & &
          Log-likelihood & $q$ & BIC & \\ \midrule
          & \multicolumn{3}{c}{Exchangeable correlation} & &
          \multicolumn{3}{c}{Unstructured correlation} \\ \midrule
$1$ & $2414.56$ & $15$ & $-4736.12$ & & $2614.84$ & $35$ & $-5012.65$ \\
$2$ & $3146.45$ & $31$ & $-6100.68$ & & $3491.12$ & $71$ & $-6542.00$ \\
$3$ & $3734.31$ & $47$ & $-7177.19$ & & $3990.00$ & $107$ & $-7316.55$ \\
$4$ & $3900.49$ & $63$ & $-7410.34$ & & $4119.53$ & $143$ & $-7352.40$ & ($\star$) \\
$5$ & $4002.73$ & $79$ & $-7515.62$ & & $4218.24$ & $179$ & $-7326.59$ \\
$6$ & $4053.61$ & $95$ & $-7518.17$ &($\star$$\star$) & $4267.97$ & $215$ & $-7202.83$ \\
$7$ & $4073.36$ & $111$ & $-7458.47$ & & $4270.47$ & $251$ & $-6984.62$ \\
$8$ & $4146.57$ & $127$ & $-7505.68$ & & $4365.99$ & $287$ & $-6952.43$ \\
$9$ & $4159.01$ & $143$ & $-7431.35$ & & $4387.43$ & $323$ & $-6772.09$ \\ \bottomrule
       \end{tabular}
      \end{small}
    \end{center}
        \caption{Maximum likelihood fits of the density (\ref{mixdensityBeta})
      for the data on the NBA players with $k \in \{2, \ldots, 9\}$
      components for both unstructured and exchangeable
      correlation structure. A ($\star$) denotes the best BIC for each
      copula specification and a ($\star\star$) the best BIC overall.}

    \label{mixBetaResults}
  \end{table}



  The maximum of the weights $w_{i1}, \ldots, w_{ik}$ at the last
  iteration of the EM algorithm (see step \emph{E1} in
  \ref{EMsection}) is used to determine the cluster membership of each
  observation.  For the model with exchangeable correlation matrices
  and $k = 6$, Figure~\ref{mixBetaMarginals} shows the marginal
  histograms of the observations within each cluster along with the
  marginal densities $g_{jt}(.)$ at the maximum likelihood estimates
  for each cluster-variable combination. The agreement of the fitted
  marginals with the histograms of the variables indicates a good
  fit. Furthermore, Figure~\ref{mixBetaMarginals} shows that the fit
  has achieved some separation between the clusters, especially for
  ``PointsHun''.

  \begin{figure}[t!]
    \begin{center}
      \includegraphics[width = 0.99\textwidth]{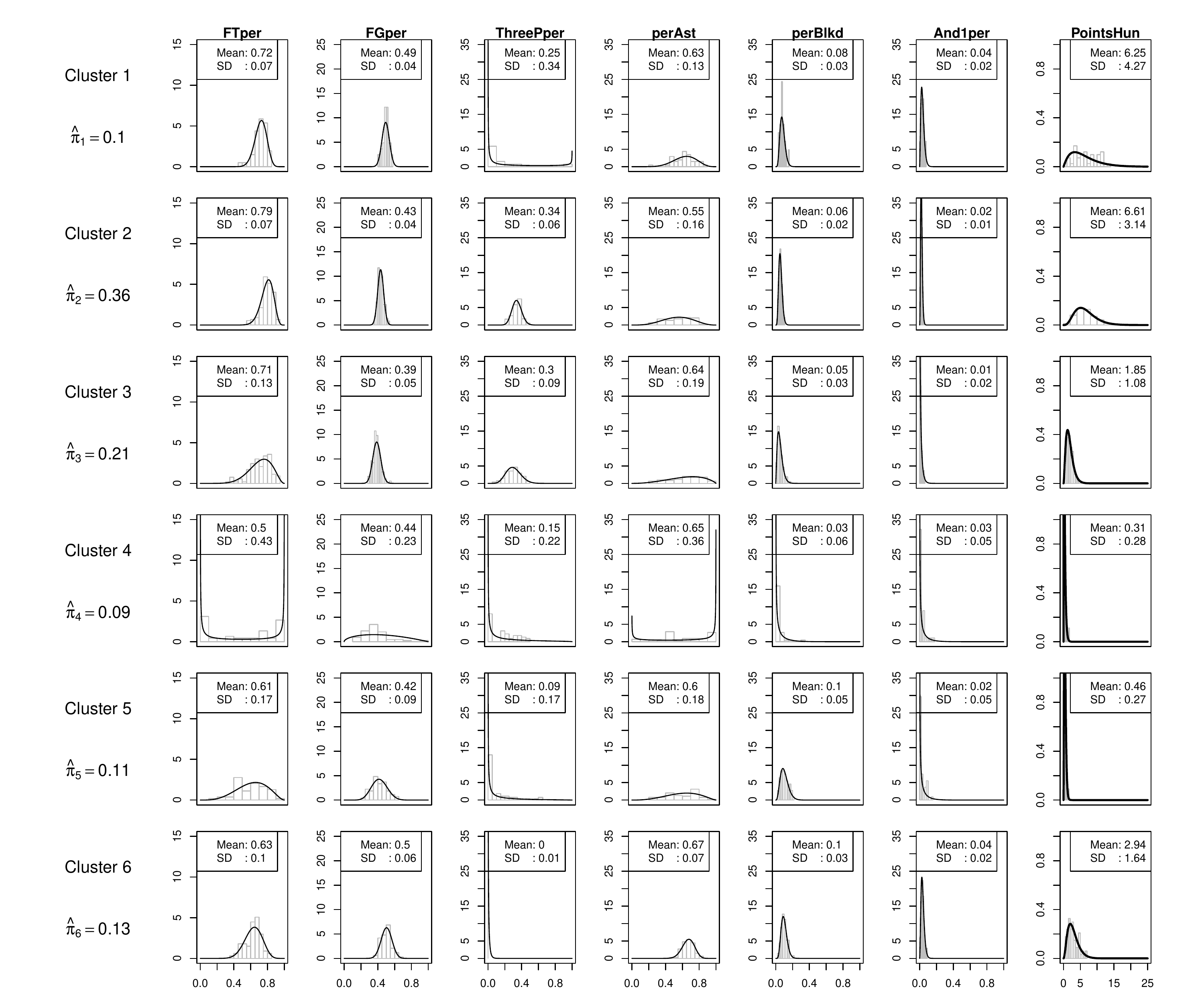}
    \end{center}
        \caption{Mixture of $6$ $7$-dimensional Gaussian copulas with
      exchangeable correlation matrices and $6$ Beta and $1$ Gamma
      marginals each. The plots show the marginal histograms of the
      observations within each cluster along with the fitted marginal
      densities for each cluster-variable combination. The fitted
      mixing proportions $\hat{\pi}_i$ $(i = 1, \ldots, 6)$ are also
      reported.}
    \label{mixBetaMarginals}
  \end{figure}

  A ``true classification'' of the players in terms of performance is
  not generally available, so it is hard to check how good the
  resultant classification is. However, there is a wealth of metrics
  that attempt to capture different characteristics of the player. A
  few representative metrics include the NBA Efficiency rating
  (``EFF''), the Usage Rate (``USG''), the True Shooting percentage
  (``TSper''), John Hollinger's Player Efficiency Rating (``PER'') and
  Alternative Win Score (``AWS''). \texttt{Hoopdata.com} provides the
  values for these metrics for each player in the 2011-2012 season and
  we use these values to assess the comparative performance of the
  copula-based mixture model to that of a Normal mixture fitted using
  the \texttt{mclust} \citep{mclust:12} R package as follows: each
  metric is broken into $I$ intervals, whose endpoints are calculated
  using the empirical quantiles at $I + 1$ equidistant probabilities
  ranging from $0$ to $1$. For each metric and for
  $I \in \{2, \ldots, 30\}$, we calculate the ARI of the clustering
  with 6 Beta and 1 Gamma marginal, and the ARI of the clustering from
  the Normal mixture model with the lowest BIC ($4$ components with
  VEV parameterization with BIC $-5026.44$; VEV stands for ``variable
  volume, equal shape, variable orientation'' and characterizes a
  particular parameterization for the variance-covariance matrix of
  the multivariate Normal distribution for the component
  distributions; see, \citealt{mclust:12} for the VEV and the other
  parameterizations that \texttt{mclust}
  uses). Figure~\ref{mixBetaValid} shows the results for each
  metric. Despite the low ARI's for both fits, all points fall below
  the $45^o$ line and, hence, the clustering from the copula-based
  mixture model clearly dominates the one from the optimal Normal
  mixture model with regards to those metrics.
    \begin{figure}[t!]
    \begin{center}
      \includegraphics[width = 0.99\textwidth]{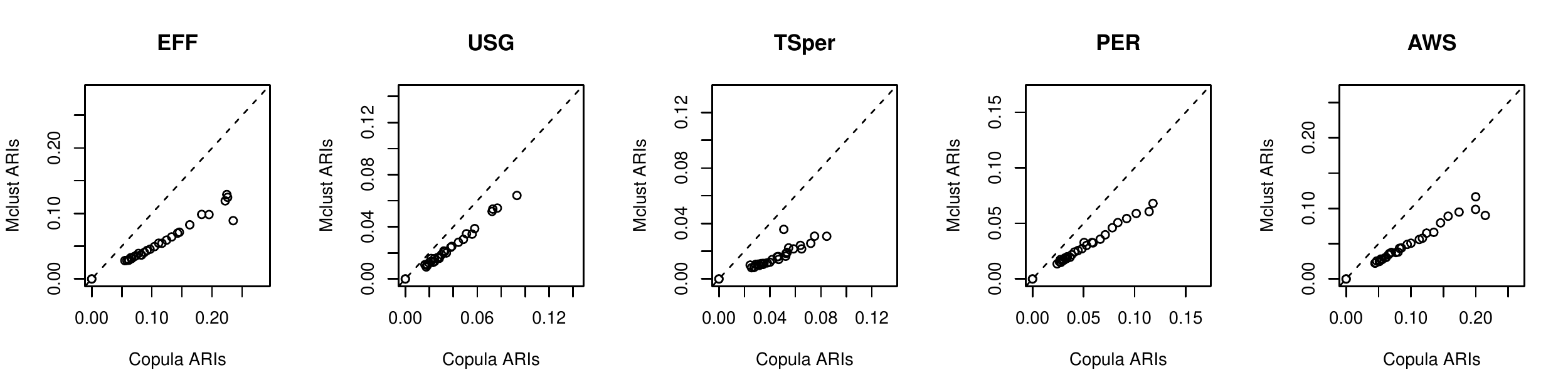}
    \end{center}
    \caption{Clustering quality of best copula-based mixture model
      versus the best Normal mixture model. The dashed line is the
      $45^o$ line from the origin.}
    \label{mixBetaValid}
  \end{figure}

\end{example}

As a reviewer correctly pointed out, another way to handle
clustering of bounded- or mixed-domain data is to first
appropriately transform the data and then use standard mixture
models for continuous data, such as Normal mixtures.
\citet{dean:13} consider such an approach by taking the arcsine
transformation of data in $[0,1]$ and then fitting Normal mixture
models on the transformed data. Their results indicate that
treating the bounded-domain data with distributions that are
defined on that domain produces better results \citep[see, for
example][Table~1, for the results of the simulation
study]{dean:13}. Another important reason for working with the
bounded-domain data directly is for avoiding the arbitrariness of
choice of transformation before fitting a mixture model. Section~1
of the supplementary material extends Example~\ref{nba} to
illustrate that distinct sensible, transformations can lead to
different results.

\subsection{Real-valued variables}
\label{realvalued}
\subsubsection{Invariance with respect to affine transformations}
\label{equivariance}
Suppose that the $n$ vectors ${\bb x}_1, \ldots, {\bb x_n}$ of
observations have real-valued components. Then, the fit of a
copula-based mixture model with marginal distributions in the
location-scale family (like Normal, skew-Normal, Cauchy, t, logistic
and so on) is invariant with respect to the translation and
component-wise scaling of ${\bb x}_1, \ldots, {\bb x}_n$. This is
because the location and scale of the components are determined only
by the marginal distributions.

More formally, suppose that all marginals support the real line
and that the fitted marginal means and variances of the $j$th
component based on data ${\bb x}_1, \ldots, {\bb x}_n$ are
$\hat{\mu}_{j1}, \ldots, \hat{\mu}_{jp}$ and $\hat{\sigma}^2_{j1},
\ldots, \hat{\sigma}^2_{jp}$ $(j = 1, \ldots, k)$, respectively.
Then, if ${\bb z}_i = a + B{\bb x}_i$ $(i = 1, \ldots, n)$, where
$B$ is a diagonal matrix with non-zero diagonal entries, the
fitted marginal means and variances of the $j$th component based
on data ${\bb z}_1, \ldots, {\bb z}_n$ will simply be $a +
B_{11}\hat{\mu}_{j1}, \ldots, a + B_{pp}\hat{\mu}_{jp}$ and
$B_{11}^2\hat{\sigma}^2_{j1}, \ldots ,
B_{pp}^2\hat{\sigma}^2_{jp}$ $(j = 1, \ldots, k)$, respectively.
This follows directly from the invariance properties of the
maximum likelihood estimator.

However, depending on the choice of copulas, the same mixture will
generally produce a different clustering for general affine
transformations of the form $z^i = a + Bx^i$ $(i = 1, \ldots, n)$,
where $B$ is a general real-valued $p \times p$ matrix. This is
because a copula-defined distribution is not necessarily closed under
general affine transformations (for example rotations). In contrast
closure under general affine transformations is satisfied for all
mixture models that are based on elliptical distributions
such as Normal and $t$ mixtures \citep{fang:02}.

\subsubsection{Rotated copulas}
\label{goodfits}
In two dimensions, the survival version of any copula $C(u_1,
u_2)$ is $C_{180}(u_1, u_2) = u_1 + u_2 - 1 + C(u_1, u_2)$, where
``180'' denotes that the survival copula is a rotated version of
$C(u_1, u_2)$ by $180^o$ clockwise. Similarly, counter-clockwise
rotated versions of $C(u_1, u_2)$ by $90^o$ and $270^o$ can be
defined as functions of $C(u_1, u_2)$ and $u_1$, $u_2$. Expression
for those are given in \citet{brechmann:12}. Hence, one can select
an initial dictionary of copulas to be used and then to further
enrich this dictionary with the rotated versions of the copulas.

\begin{example}
  \label{survival}
  In Example~\ref{motivatingExampleCont} we chose the copulas for the
  components by recognising the need for mixture components that can
  accommodate extreme tail dependence through inspection of the
  scatterplot of the data. In this respect, we chose to have two
  mixture components based on the Gumbel copula which exhibits
  upper tail dependence and two components based on the Clayton copula
  which exhibits lower tail-dependence. Since a rotated version of
  the Clayton copula by $180^o$ (survival Clayton) would exhibit
  upper tail dependence one might argue that a mixture model that uses
  the rotated version of the Clayton instead of Gumbel would have
  produced a fit of comparable quality. Indeed that is the case;
  fitting a model with two Clayton and two survival Clayton components
  gives misclassification error of $9.875\%$ and a BIC value of
  $5561.73$. Similarly, fitting a model with two Gumbel and two
  survival Gumbel components gives misclassification error of
  $11.25\%$ and a BIC value of $5644.36$. \qed
\end{example}

As is illustrated in Example~\ref{survival}, the ability to
construct new copula families from known ones certainly adds great
flexibility when constructing copula-based mixture models for
clustering. However, it certainly does not simplify model
selection in a mixture model framework; if we limit ourselves to a
dictionary of $d$ copulas, then for finding the best model amongst
the models with $k$ components we need to fit and select the best
model from $\binom{d + k - 1}{k}$ models. Furthermore, if the
number of components is also considered as part of the model
selection exercise, then one needs to fit and compare $\sum_{k =
1}^K\binom{d + k - 1}{k}$ where $K$ is a preset maximum number of
components. Both these numbers increase quickly as either $K$ or
$d$ increase possibly making the model selection exercise
impractical.

\subsubsection{Component-wise parametric rotations}
Subsections~\ref{equivariance} and \ref{goodfits} show that the
added flexibility from the use of mixture of copulas may come with
the price of two shortcomings from a practitioners point of view:
the general lack of invariance with respect to general affine
transformations of the data, and the fact that completely
different copulas can result in fits of comparable quality. The
latter issue is not so serious provided that the computational
resources are enough for fitting many models and keeping in mind
the target of the analysis is to find a good model. Though it
certainly points towards the direction that if one had a more
flexible specification for the mixture components, the number
$\sum_{k = 1}^K\binom{d + k - 1}{k}$ of models that need to be
fitted is significantly decreased because $d$ can be drastically
reduced. The invariance issue is harder to tackle. However, a
flexible enough model could alleviate some of the invariance
issues by maintaining the translation and scaling invariance of
the component densities, and by allowing the component densities
to rotate based on the observations.

Consider the mixture of copulas specified by (\ref{mixture_model}) and
(\ref{basic_model}) with $p = 2$. Temporarily omitting the component
index, each component density has the form
\[
f({\bb x}^*; {\bb \gamma}, {\bb \psi}) = c(G_1(x_1^*; {\bb
\gamma}_1), G_2(x_2^*; {\bb \gamma}_2); {\bb \psi})g_1(x_1^*;{\bb
  \gamma}_1)g_2(x_2^*;{\bb \gamma}_2) \, , \quad {\bb x}^* \in \Re^2\,
.
\]
Now consider the transformation ${\bb X} = O(\omega) {\bb X}^*$, where
$O(\omega)$ is the rotation matrix
\[
O(\omega) = \left[
  \begin{array}{cc}
    \cos \omega & -\sin\omega \\
    \sin\omega & \cos \omega \\
  \end{array}
\right] \, ,
\]
with $\omega \in (0, 2\pi]$. Then ${\bb X}$ is a counter-clockwise
rotation of ${\bb X}^*$ at an angle $\omega$ and the density function
of ${\bb X}$ is simply
\begin{equation}
  \label{rotatedComponent}
  f^*({\bb x}; {\bb \gamma}, {\bb \psi}, \omega) = f(O(\omega)^\top{\bb
    x}; {\bb \gamma}, {\bb \psi})\, , \quad
  {\bb x} \in \Re^2 \, ,
\end{equation}
because for any rotation matrix $O(\omega)^{-1} = O(\omega)^\top$ ($O(\omega)$ is
orthonormal) and $|\det O(\omega)| = 1$. Hence, the contours of $f^*$
will be a counter-clockwise rotation of the contours of $f$ at an
angle $\omega$. Hence, in the two-dimensional case, an extended
mixture model can be defined that has the form
\begin{equation}
  \label{rotatedMixture}
  h({\bb x}; {\bb \theta}, {\bb \omega}) = \sum_{j = 1}^k \pi_j
  f^*_j({\bb x}; {\bb \gamma}_j, {\bb \psi}_j, {\bb \omega}_j) \, .
\end{equation}
The difference of the latter specification from the mixture model in
(\ref{mixture_model}) and (\ref{basic_model}) is that there are an
extra $k$ rotation angles to be estimated, but the added flexibility
is enormous. The practitioner can now select the marginals and a much
smaller dictionary of copulas; notice that all versions of the rotated
copulas by $90^o$, $180^o$ and $270^o$ are special cases for the
components of model~(\ref{rotatedMixture}) for specific values of
$\omega_1, \ldots, \omega_k$ and that other exotic bivariate
distributions may result for arbitrary angles.

The mixture model (\ref{rotatedMixture}) can be fitted using the
EM algorithm; the only modification from the general iteration in
Subsection~\ref{EMsection} is that at {\em M-step 2} of the
$\ell$th iteration, the function
\begin{equation}
  \sum_{j = 1}^k \sum_{i = 1}^n
  w_{ij}^{(\ell + 1)} \left[ \log c_j(G_1(z_{i1}(\omega_j); {\bb \gamma}_{j1}), \ldots,
    G_p(z_{ip}(\omega_j); {\bb \gamma}_{jp}); {\bb \psi}_j) +
    \sum_{t = 1}^p\log g_t(z_{it}(\omega_j); {\bb \gamma}_{jt})  \right] \, ,
    \label{rotated}
\end{equation}
is maximized with respect to ${\bb \gamma}_{11}, \ldots, {\bb
  \gamma}_{1p}, \ldots, {\bb \gamma}_{k1}, \ldots, {\bb \gamma}_{kp},
{\bb \psi}_1, \ldots, {\bb \psi}_k, \omega_1, \ldots, \omega_k$.
In (\ref{rotated}), ${\bb z}_i(\omega_j) = O(\omega_j)^\top{\bb
x}_i$ where ${\bb z}_i(\omega_j) = (z_{i1}(\omega_j), \ldots,
z_{ip}(\omega_j))^\top$ $(i = 1, \ldots, n; j = 1, \ldots, k)$.
This {\em M-step} can also be broken down into $k$ independent
optimizations.

In order to avoid maximization over a large parameter space, the ECM
algorithm in Subsection~\ref{maxContinuous} can be extended for
handling parametric rotations i) by replacing $x_{it}$ with
$z_{it}(\omega_j^{(\ell)})$ $(t = 1, \ldots, p)$ in {\em CM-step 1}
and {\em CM-step 2}, and ii) by including an additional last step to
update the angles at the values of the marginal and copula parameters
from {\em CM-step 1} and {\em CM-step 2}. In that last step the
objective
\begin{align}
  \label{CM3continuous}
  \sum_{j = 1}^k \sum_{i = 1}^n w_{ij}^{(\ell + 1)} & \Bigg[ \log
    c_j(G_1(z_{i1}(\omega_j); {\bb \gamma}^{(\ell + 1)}_{j1}), \ldots,
    G_p(z_{ip}(\omega_j); {\bb \gamma}^{(\ell + 1)}_{jp}); {\bb
      \psi}^{(\ell + 1)}_j) \\ \notag
  + & \sum_{t = 1}^p\log g_t(z_{it}(\omega_j);
    {\bb \gamma}^{(\ell + 1)}_{jt}) \Bigg] \, ,
\end{align}
is maximized with respect to $\omega_1, \ldots,
\omega_k$ to obtain updated values $\omega_1^{(\ell + 1)}, \ldots,
\omega_k^{(\ell +
  1)}$. As is the case for {\em CM-step 1} and {\em CM-step 2}, this
last step can also be broken down into parallel optimizations
across components, each of which consists of a one-dimensional
maximization with respect to the respective angle.

\begin{example}
  \label{rotexample}
  As is illustrated in Example~\ref{survival} one can fit a mixture of
  two Clayton and two survival Clayton, or a mixture of two Gumbel and
  two survival Gumbel, or a mixture of two Gumbel and two Clayton
  copulas to the artificial data of Example~\ref{motivatingExample}
  and obtain comparable fits in all cases. Hence, we have considered the
  combinations of four different copulas for the components so
  far. The whole modelling exercise is much easier if we pick just one
  copula which exhibits tail-dependence (upper or lower), and
  Normal marginals and use those for setting up the mixture density
  (\ref{rotatedMixture}).

  For example, using the Clayton copula one obtains the estimated angles
  $\hat{\omega}_1 = 180.41$, $\hat{\omega}_2 = 180.78$,
  $\hat{\omega}_3 = 7.13$, $\hat{\omega}_4 = 359.99$ and a
  misclassification error of
  $9.875\%$. Figure~\ref{artificialRotations} shows the contours of
  the fitted component densities across the iterations of the ECM
  algorithm and demonstrates the enormous flexibility that parametric
  rotations offer when setting copula-based mixture models. Iteration
  0 (top left) refers to the starting values for the ECM algorithm.
  \begin{figure}[t!]
    \begin{center}
      \includegraphics[width =
      0.99\textwidth]{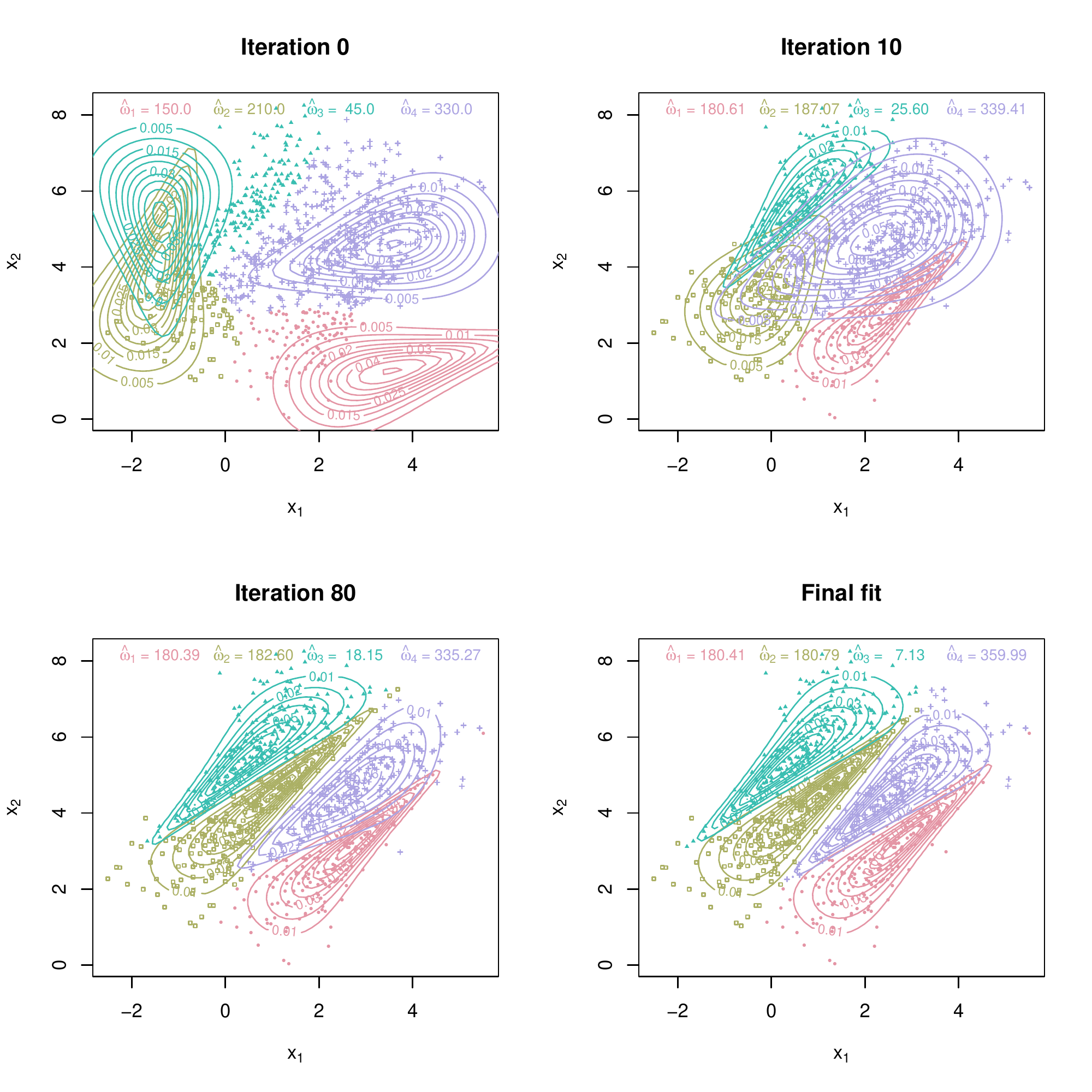}
    \end{center}
        \caption{The contours of the fitted component densities across the
      iterations of the ECM algorithm when allowing for parametric
      rotations. The value of the component angles at the depicted
      iterations is given in the top of each plot. Iteration 0 (top
      left) refers to the starting values for the ECM algorithm.}

    \label{artificialRotations}
  \end{figure}
  The BIC value for the fit that allows parametric rotations is
  $5585.372$ which is larger the $5561.73$ of the mixture with two
  Clayton and two survival Clayton in Example~\ref{survival}. This
  slight inflation is due to the 4 extra parameters included in the
  model that allows for component-wise parametric rotations. \qed
\end{example}


\subsubsection{Identifiability of rotations for elliptical distributions}
It should be noted that when at least one of the component
distributions is elliptical then the corresponding angles in
(\ref{rotatedMixture}) are not identifiable. To show that, suppose
that ${\bb X}^*$ has density $f({\bb x}^*;{\bb \gamma},{\bb
\psi})$ with mean ${\bb \mu}^*$ and variance-covariance matrix
$\Sigma^*$. If ${\bb X}$ has density (\ref{rotatedComponent}), the
variance-covariance matrix of ${\bb X}$ is
\begin{align*}
  \notag {\var}({\bb X}) & = E_{f^*}\left[\left\{{\bb X} - {\bb
        \mu}(\omega)\right\}\left\{{\bb X} - {\bb
        \mu}(\omega)\right\}^\top\right] \\
  \notag & = E_{f}\left\{O(\omega)({\bb X}^* - {\bb \mu}^*)({\bb X}^*
    -
    {\bb \mu}^*)^\top O(\omega)^\top\right\}  \\
  & = O(\omega)\Sigma^* O(\omega)^\top \, .
\end{align*}
The variance-covariance matrix $\Sigma^*$ admits the
eigen-decomposition $\Sigma^* = Q_j \Lambda_j Q_j$ where $Q_j$ is
an orthogonal matrix. Noting that the product of two-orthogonal
matrices is orthogonal $O(\omega)\Sigma^* O(\omega)$ is also a
variance covariance-matrix and given that we can only identify
variances and covariances, the angle $\omega$ is not identifiable.
For example, if Normal marginals are considered, then using a
Gaussian copula for one of the components of
(\ref{rotatedMixture}) will result to identifiability issues.

\subsubsection{Local rotation unidentifiability for general copulas}
Furthermore, as is discussed in \citet[Section~4.3]{nelsen:06},
many Archimedean copulas have the product copula $C(u, v) = uv$ as
a special case for specific values of their parameters (for
example, the Clayton copula for $\psi \to 0$, the Gumbel for $\psi
\to 1$, the Frank for $\psi \to 0$, and so on). The use of
parametric rotations poses local identifiability problems, for
those specific boundary values of the copula parameter.

\section{Closure under marginalization}
\label{closure} A desirable property that well-used mixture models
such as mixtures of multivariate Normal, multivariate skew-Normal
and multivariate skew-t distributions share is closure under
marginalization. Such a property guarantees that the marginal of
any dimension of the component distributions belongs to the same
family of distributions as the component distribution itself, and
allows the easy transition from the full mixture model to a
marginal model of any order. For example, one can fit a mixture of
multivariate Normal distributions and then plot the contours of
all bivariate marginal densities by using the bivariate Normal
density and the appropriate subsets of parameters from the full
model without the need of integrating over the fitted density.

For $m < p$ and continuous random variables the requirement of closure
under marginalization for the component density is that if
\[
f_j^{(m)}(x_1, \ldots, x_{m}; \theta_j) =
\int_{\mathcal{X}_{m+1}}\ldots \int_{\mathcal{X}_p} f_j(x_1, \ldots,
x_p; \theta_j) d x_{m+1} \ldots d x_{p} \quad (j = 1, \ldots, k) \, ,
\]
where $\mathcal{X}_{m+1}, \ldots, \mathcal{X}_p$ are the supports of
the random variables $X_{m+1}, \ldots, X_p$, respectively, then
$f_j^{(m)}$ has exactly the same functional form as the $p$
dimensional density $f_j$ does but in $m$ dimensions. For a
copula-defined $p$-dimensional component density the requirement from
the copula distribution is that, if
\[
C_j^{(m)}(G_1(x_1; \gamma_{j1}), \ldots, G_m(x_m; \gamma_{jm});
\psi_j) = C_j(G_1(x_1; \gamma_{j1}), \ldots, G_m(x_{m}; \gamma_{jm}),
1, \ldots, 1; \psi_j) \, ,
\]
then $C_j^{(m)}$ belongs to the same family of copulas as $C_j$
does. To derive this requirement, marginalization is performed by
setting $x_{m+1}, \ldots, x_{p}$ to the maximum of their range of
definition. This property is satisfied for all elliptical copulas
like the Gaussian copula and the $t$-copula \citep[see][for
results on the family of elliptical copulas]{fang:02}.
Furthermore, closure under marginalization holds for every
Archimedean and nested Archimedean copula, because its generator
function necessarily takes value 0 at 1 \citep[see, for
example,][for definitions and results for multivariate Archimedean
and nested Archimedean copulas]{hofert:12}.

The results here extend to the case of discrete data by replacing the
integration of density functions with summations of probability mass
functions of the form (\ref{copDiscrete}). Closure under
marginalization is a particularly relevant property to be satisfied
when building mixture models for discrete data, because, otherwise,
the computational burden involved in the calculation of marginals of
the mixture model can be prohibitive.

A class of copulas that does not satisfy the property of closure
under marginalisation is the class of vine copulas (see, for
example, \citealt{bedford:02}).

\begin{example}
  In Example~\ref{nba}, the mixture components were defined using the
  Gaussian copula. Hence, the bivariate marginal density of $X_t$ and
  $X_s$ $(s,t = 1,\ldots,7; s\ne t)$ corresponding to the density
  (\ref{mixdensityBeta}), are simply
  \begin{equation}
    \label{mixdensityBetaMarginal}
    \sum_{j = 1}^k \pi_j \phi_2\left[ \Psi\left\{G_{js}(x_s)\right\},
      \Psi\left\{G_{jt}(x_{t})\right\}; R_{j,st} \right]
    \frac{g_{js}(x_s)g_{jt}(x_{t})}{\phi_1[\Psi\{G_{js}(x_s)\}]\phi_1[\Psi\{G_{jt}(x_t)\}]}
    \, ,
  \end{equation}
  where $R_{j, st}$ is the $2 \times 2$ correlation matrix with the $(s, s)$th and
  $(t, t)$th components of $R_j$ in the diagonal and the $(s, t)$th
  component of $R_j$ in the off-diagonal $(j =1, \ldots, k)$.

  \begin{figure}[t!]
    \begin{center}
      \includegraphics[width =
      0.99\textwidth]{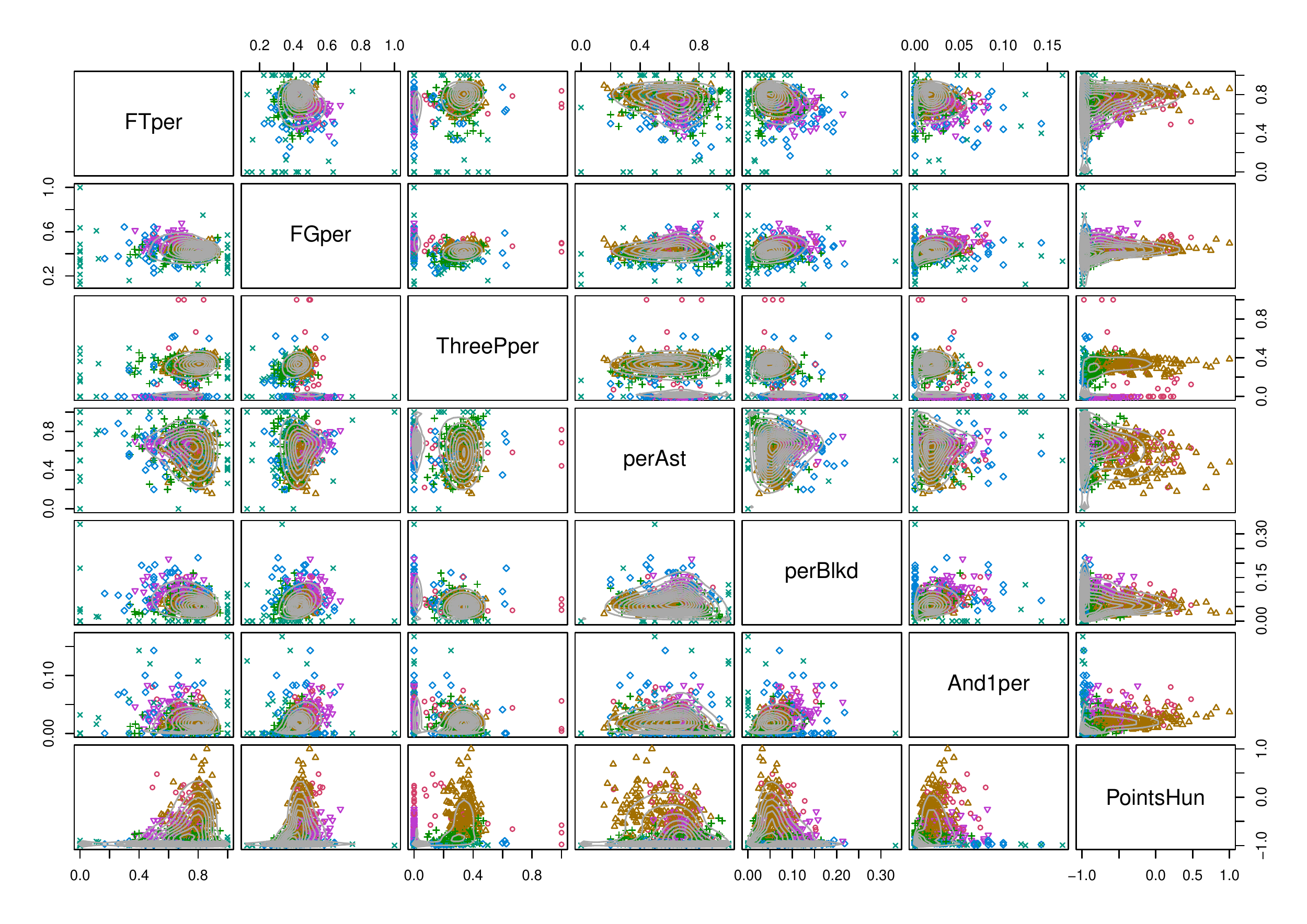}
    \end{center}
    \caption{The contours of all the bivariate marginal densities of
      the mixture model with $6$ components and unstructured
      correlation matrices from Example~\ref{nba}. The observations
      are coloured according to their assigned cluster.}
    \label{bivariateMarginals}
  \end{figure}

  Figure~\ref{bivariateMarginals} shows the contours of the bivariate
  marginals for the best model according to BIC in
  Table~\ref{mixBetaResults}, with the data coloured according to
  their assigned cluster.  \qed

\end{example}


\section{Discrete data}


\subsection{Copula-based mixture models}
The general mixture model specification and fitting framework set in
Section~\ref{mixturemodels} and Section~\ref{modelfitting} can
directly be used for constructing and estimating mixture models with
discrete marginal distributions. Nevertheless, in the discrete case,
model specification and estimation need a more careful consideration
than they do in the continuous case.

\subsection{Model specification}
The choice of the component copulas cannot be based entirely on
dependence considerations (tail dependence, correlation, and so on) as
is rather intuitively done in the continuous case. This is because the
copula alone does not anymore characterize the dependence between the
discrete marginals; the usual dependence measures, like Kendall's
$\tau$ and Spearman's $\rho$, are not margin-free as they are in the
continuous setting. \citet[\S 4]{genest2007primer} provide theoretical
derivations and demonstrations of such issues, with detailed
discussions of how they reflect in practice. Furthermore, as discussed
in Section~\ref{closure} and as is illustrated by
Example~\ref{cognitive} below, closure under marginalization is
particularly relevant for mixture models for discrete data, if
marginal assessments of the clustering or fit are to be obtained.

\subsection{Estimation}
For the analysis of discrete data, {\em M-step 2} of
Subsection~\ref{EMsection} takes the form
\begin{itemize}
\item {\em M-step 2:} Maximize the log-likelihood
  \begin{equation}
    \label{M2discrete}
    \sum_{j = 1}^k \sum_{i = 1}^n
    w_{ij}^{(\ell + 1)} \left[ \log \sum_{{\bb d}_i} \sign{{\bb d}_i}
    C(G_1(d_{i1};
      {\bb \gamma}_{j1}), \ldots, G_p(d_{ip}; {\bb \gamma}_{jp}); {\bb \psi}_j)  \right] \, ,
  \end{equation}
\end{itemize}
with respect to ${\bb \psi}_1, \ldots, {\bb \psi}_k, {\bb
  \gamma}_{11}, \ldots, {\bb \gamma}_{1p}, {\bb \gamma}_{k1}, \ldots,
{\bb \gamma}_{kp}$, where ${\bb d}_i$ is as in expression
(\ref{copDiscrete}) for an observation ${\bb x}_i$ $(t = 1, \ldots, p;
j = 1, \ldots, k; i = 1, \ldots, n)$. Note here that in the discrete
case the ECM algorithm would offer no simplification over EM, since a
copula-specified probability mass function cannot be decomposed as in
the continuous case.

The combination of $\sum_{t = 0}^p\binom{p}{t}$ summation terms in
the right-most summation in (\ref{M2discrete}) and the lack of
closed-form expression for general copulas can result in the
accumulation of numerical error that in turn can lead in
calculated probability mass functions less than $0$ or greater
than $1$ for certain parameter settings, which can result in
computational problems in the {\em M-step}.

A partial resolution of those issues, at least for small to
moderate $p$, exists for copulas that are derived from well-known
distribution functions through the inverse probability transform.
Let $C(u_1, \ldots, u_p) = H(H_1^{-1}(u_1), \ldots,
H_p^{-1}(u_p))$, where $H(., \ldots, .)$ is some $p$-variate
distribution function with marginals $H_1(.), \ldots, H_p(.)$ and
$H_j^{-1}(.)$ is the quantile function of $H_j(.)$ $(j = 1,
\ldots, p)$. Then, omitting the component index and suppressing
the dependence on the parameters,
\begin{equation}
\label{rectangle} \sum_{{\bb d}} \sign{{\bb d}} C(G_1(d_{1}),
\ldots, G_p(d_{p})) = \int_{D(x_1)}\ldots\int_{D(x_p)} h(y_1,
\ldots, y_p)dy_p \ldots dy_1\, ,
\end{equation}
where $D(x_t)$ is the interval from $H_t^{-1}\{G_t(x_t-1)\}$ to
$H_t^{-1}\{G_t(x_t)\}$ $(t = 1, \ldots, p)$, and $h(., \ldots, .)$ is
the density function corresponding to $H(., \ldots, .)$.

Hence, a single evaluation of the rectangle probability in
(\ref{rectangle}) is sufficient for calculating the probability mass
function. For special but prominent copulas like the Gaussian and the
$t$ copula and for not very large $p$, the probability mass function
can be calculated through accurate approximation methods like those of
\citet{joe:95}.
Such methods are implemented in the {\tt mprobit} R package by Joe,
Choy and Zhang and the {\tt mvtnorm} R package \citep{mvtnorm:13}. The
following example concerns the use of copulas to construct mixtures of
trivariate Binomial distributions that allow for dependence.

\begin{example}
  \label{cognitive}

  This example relates to cognitive diagnosis modelling. The data set
  consists of the responses of 536 middle school students on 20 items
  of a fraction subtraction test.
  Each item can belong to more than one attribute that one wants to
  measure. Hence, attribute scores for each student can be obtained by
  counting the number of successful items out of the total items that
  belong to each attribute. The data are available in the {\tt CDM} R
  package \citep{cdm:14} and its documentation describes which items
  belong to which attribute. The aim of this example is to use some of
  the attribute scores of the students for the construction of
  performance clusters of the latter. The scores that are used in this
  example are for the attributes ``separate a whole number from a
  fraction'' (score $X_1$), ``borrow from whole number part'' (score
  $X_2$) and ``subtract numerators'' (score $X_3$), which are traced
  on 13, 8 and 19 items, respectively. Hence, a natural distributional
  choice for each of $X_1$, $X_2$ and $X_3$ is Binomial. More
  specifically, we assume that for the $j$th component, the $t$th
  marginal distribution is Binomial with index $m_j$ and probability
  of success $p_{jt}$ $(t = 1, 2, 3; j= 1,\ldots, k)$, where $m_1 =
  13$, $m_2 = 8$ and $m_3 = 19$.

  \begin{figure}[t!]
    \begin{center}
      \includegraphics[width = 0.32\textwidth]{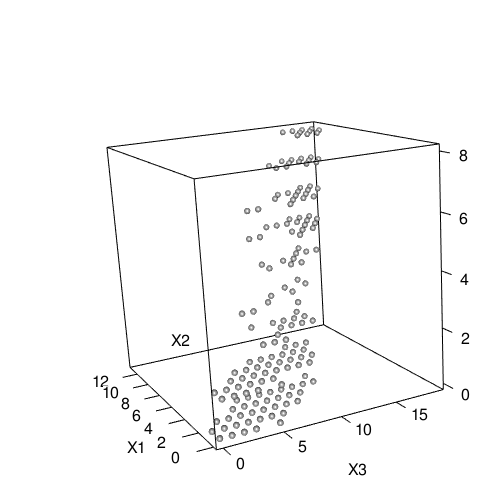}
      \includegraphics[width = 0.32\textwidth]{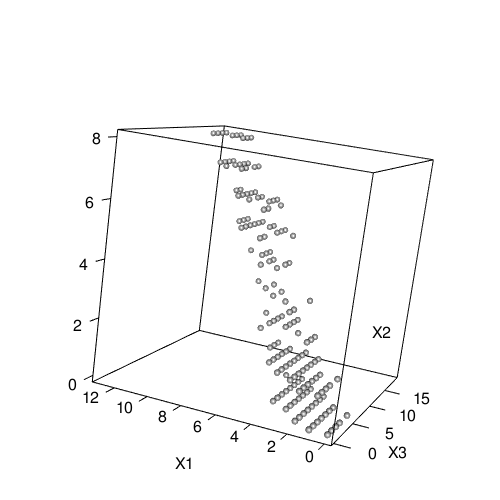}
      \includegraphics[width = 0.32\textwidth]{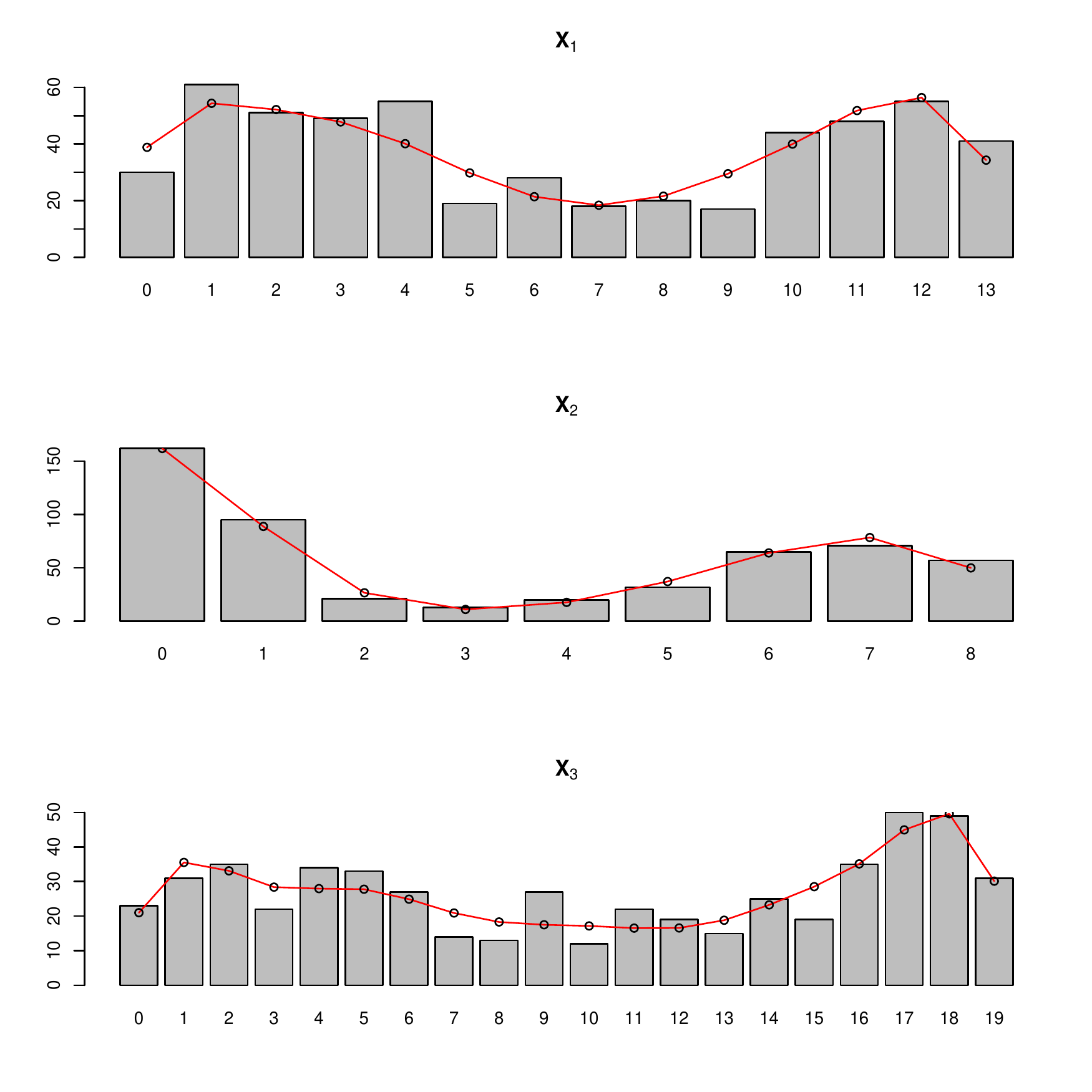}
    \end{center}
    \caption{Different views of the three attribute scores for all
      students. The barplot at the rightmost plot shows the observed
      (bars) and the fitted (points) frequencies for the three
      attributes based on the selected model.}
    \label{CognitiveData}
  \end{figure}


  Figure~\ref{CognitiveData} presents the data from two different
  angles and reveals that there is strong positive association between
  each pair of attribute scores. This is because the three attributes
  share items, and this association needs to be taken into account
  when clustering the students.  The barplot at the rightmost plot shows the observed
      (bars) and the fitted (points) frequencies for the three
      attributes based on the selected model. Such data have also been analysed in
  the past via mixture models \citep[see,][]{dean:13} but only after
  transforming the scores into percentages and treating those as
  realizations of continuous random variables. Such transformations
  are not necessary when using copula-based mixture models; to
  accommodate for the apparent association, we construct $4$ different
  families of mixture models each with components that are trivariate
  Binomial distributions defined using i) one-parameter Frank copulas,
  ii) two-parameter Frank copulas \citep[see,][]{zimmer:06}, iii)
  trivariate Gaussian copulas as in (\ref{multivariateNormal}) with
  unstructured correlation (one parameter each), and iv) trivariate
  Gaussian copulas with exchangeable correlation (three parameters
  each). Note here that defining multivariate Binomial distributions
  which allow for correlated marginals is not straightforward outside
  the copula framework \citep[for a discussion on bivariate and
  multivariate Binomial distributions, see][]{johnson:97}.

  \begin{table}
    \begin{center}
      \begin{tabular}{cccclcccl}
        \toprule
        &\multicolumn{3}{c}{One-parameter
          Frank}&&\multicolumn{3}{c}{Gaussian - exchangeable
          correlation}\\ \midrule
$k$&Log-likelihood&$q$&BIC&&Log-likelihood&$q$&BIC\\ \midrule
1&-5899.71&4&11824.56&&-5063.32&4&10151.78\\
2&-3098.11&9&6252.78&&-3049.76&9&6156.08\\
3&-2843.22&14&5774.42&&-2846.08&14&5780.14\\
4&-2758.64&19&5636.68&&-2750.57&19&5620.54\\
5&-2720.35&24&5591.52&&-2707.63&24&5566.08\\
6&-2693.26&29&5568.76 & ($\star$) &-2679.56&29&5541.36 &($\star$$\star$) \\
7&-2678.67&34&5570.99&&-2669.90&34&5553.45\\
8&-2666.62&39&5578.32&&-2662.46&39&5570.00\\
 \midrule
        &\multicolumn{3}{c}{Two-parameter
          Frank}&&\multicolumn{3}{c}{Gaussian - unstructured
          correlation}\\ \midrule
        $k$&Log-likelihood&$q$&BIC&&Log-likelihood&$q$&BIC\\ \midrule
1& -5897.92&5&11827.261&&-5010.11&6&10057.925\\
2&-3085.49&11&6240.11&&-2990.15&13&6061.99\\
3&-2825.91&17&5758.65&&-2773.45&20&5672.58\\
4&-2750.89&23&5646.32&&-2709.79&27&5589.26\\
5&-2711.81&29&5605.86&&-2674.92&34&5563.50 &($\star$) \\
6&-2673.79&35&5567.52& ($\star$) &-2671.35&41&5600.35\\
7&-2659.98&41&5577.60&&-2662.37&48&5626.38\\
8&-2654.7&47&5604.76&&-2658.42&55&5662.47\\
        \bottomrule
      \end{tabular}
      \caption{\label{Table1} Results from fitting finite mixtures
        with different number of components. A ($\star$) denotes the
        best BIC for each copula specification and a ($\star\star$)
        the best BIC overall.}
    \end{center}
  \end{table}

  The one-parameter Frank copula is defined as
  \begin{equation}
    \label{Frank}
    C^{(F)}(u_1,u_2,u_3;
    \psi)=-\frac{1}{\psi}\log\left\{1+\frac{(\exp^{-\psi
          u_1}-1)(\exp^{-\psi u_2}-1)(\exp^{-\psi
          u_3}-1)}{(\exp^{-\psi}-1)^2}\right\}\, ,
  \end{equation}
  where $\psi$ is an association parameter which is common for all
  marginals, implying symmetric dependence. The two-parameter Frank
  copula is a trivariate nested Archimedean copula which has been used
  in the applications in \citet{zimmer:06} and is defined as
  \begin{equation}
    \label{2parFrank}
    C^{(F^*)}(u_1,u_2,u_3; {\bb\psi})= \\
    -\frac{1}{\psi_1} \log\left[ 1-\frac{1}{\gamma_1}
      \left\{1-\big(1-\frac{1}{\gamma_2} \tau_1 \tau_2
        \big)^{\psi_1/\psi_2} \right\}(1-\exp(-\psi_1 u_3)) \right]\, ,
  \end{equation}
  where $0<\psi_1<\psi_2$, $\zeta_1=1-\exp(-\psi_1)$,
  $\zeta_2=1-\exp(-\psi_2)$, $\tau_1 =1-\exp(-\psi_2 u_1) $ and
  $\tau_2 =1-\exp(-\psi_2 u_2)$. The two-parameter Frank copula has
  (\ref{Frank}) as a special case, and with one extra parameter it
  allows for partial symmetry capturing more flexible associations
  between the marginals.

Since (\ref{Frank}) and (\ref{2parFrank}) are of closed-form, the
  calculation of the probability mass function for the components of
  the respective mixture models is performed using
  (\ref{copDiscrete}), which in the current case of $3$ variables
  consists of $8$ terms.  For the multivariate Gaussian copula, the
  probability mass functions for the components are instead obtained
  via the approximation of the rectangle probability in
  (\ref{rectangle}) using the {\tt mprobit} R package. In order to
  avoid overflows due to the approximation of the multivariate Normal
  integral, probabilities calculated as smaller than $10^{-12}$ were
  kept to this value.

  \begin{figure}[t!]
    \begin{center}
      \includegraphics[width = 0.60\textwidth]{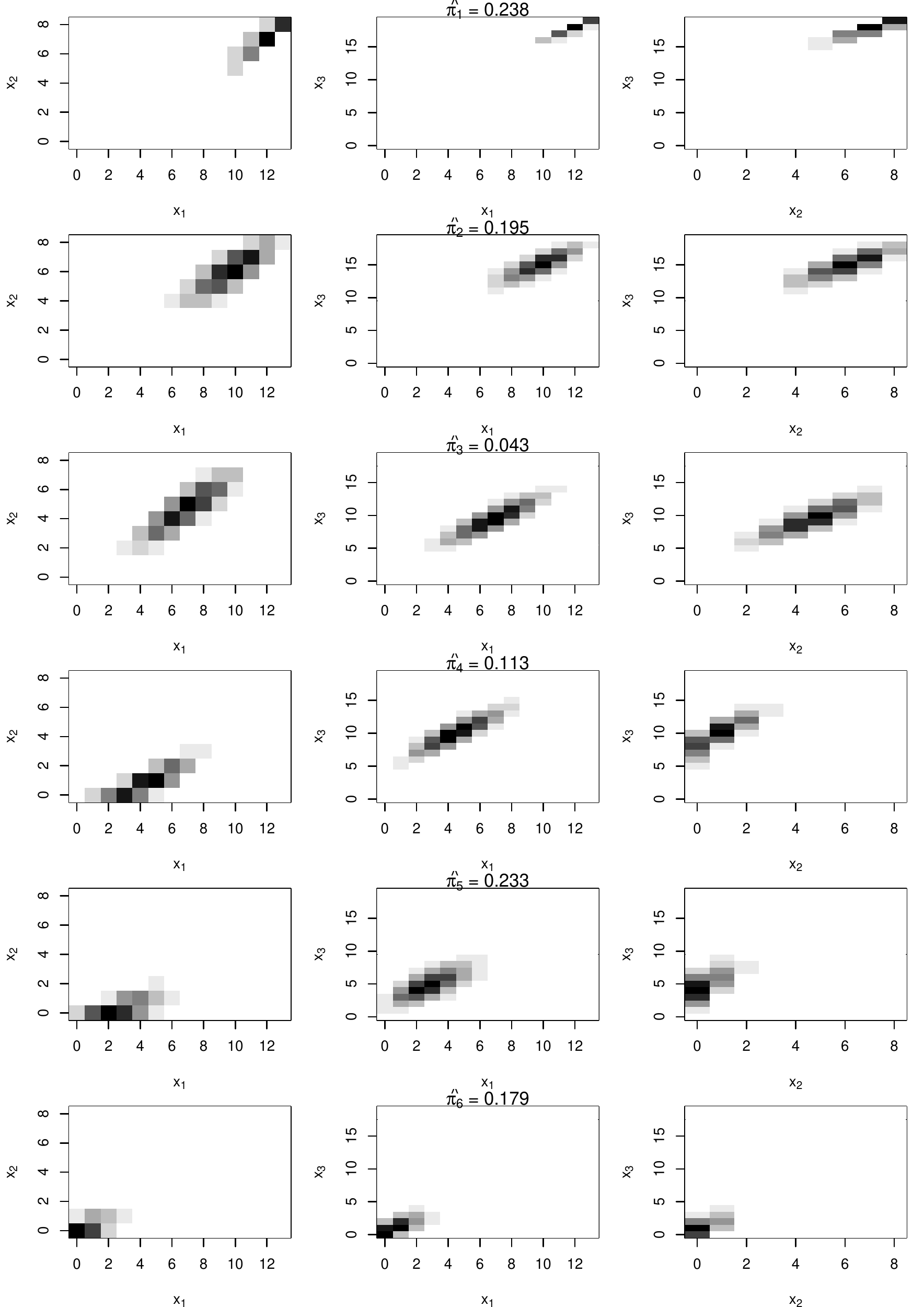}
    \end{center}
      \caption{Fitted components from the model with the selected
        model with Gaussian copula with exchangeable correlation and 6
        components.}
      \label{figbin}
  \end{figure}

  The finite mixture given in (\ref{mixture_model}) was fitted using
  the EM algorithm of Section~\ref{EMsection}, where now ${\bb
    \theta}_j = (\bb{\psi}_j, p_{j1}, p_{j2}, p_{j3})$ $(j =1 ,
  \ldots, k)$.  Starting values were obtained using the approach
  described in Subsection~\ref{starting} with 10 sets of random
  starting values. In parallel to that the following sequential
  approach has been applied: for $k = 1$, estimates for the Binomial
  proportions were obtained by the marginals, while the correlation
  parameters were set equal to the sample counterparts or their
  average in the cases when one correlation parameter is used for all
  pairs. For the two-parameter Frank copula, one parameter was set equal to
  the smallest of the three correlations and the other to the same
  value plus a small positive number in order to satisfy the
  restriction $\psi_1<\psi_2$. Then starting values for the model with
  $k + 1$ components were obtained by using the parameters of the
  model with $k$ components and by adding a new component with parameter values those
  found when fitting a model with just one component. This new component was
  given a small mixing proportion (we used 0.05). This procedure
  worked well and provided the largest maximized log-likelihood for
  almost all $k$.

  Table \ref{Table1} shows the results from fitting a series of
  models. The maximized log-likelihood, the number of parameters $q$
  and the value of BIC are reported for each model. The overall best
  model according to BIC is noted with ($\star\star$) and is the model
  with 6 components specified using Gaussian copulas with exchangeable
  correlation.

  The Gaussian copula is closed under marginalisation and this allows
  the effortless calculation of bivariate marginals (see
  Section~\ref{closure} for details).  Each row of Figure~\ref{figbin}
  corresponds to one of the fitted trivariate Binomial components, and
  shows the $3$ possible bivariate marginals of that component. The
  label of each row gives the corresponding mixing proportion. The
  darker the color on each plot the larger is the probability mass for
  the corresponding combination of values of the Binomial
  variables. The component in the first row corresponds to the
  students that scored very well in the test in all attributes, and
  the component in the last to those with the worst results.  As is
  apparent in the plots of Figure~\ref{figbin}, there is a strong
  positive correlation in most of the components, which implies a
  general ability of different levels in all three attributes.  A few
  components deviate from this pattern. For example, the component in
  the fourth row from the top corresponds to students that scored
  poorly in $X_1$ and $X_2$ but moderately in $X_3$. Also, the
  component in the fifth row from the top corresponds to students
  whose scores for $X_1$ and $X_2$ are very close to $0$ and the
  scores for $X_3$ are slightly higher. Lastly, the barplot in Figure
  \ref{CognitiveData} shows that the selected model fits
  satisfactorily the observed frequencies, despite the rather
  complicated behaviour they demonstrate.
\end{example}


\section{Discussion and further work}
\subsection{Advantages}
This paper introduces a general framework for model-based clustering
where the component densities can be specified through copulas.The
numerous examples in this paper both on real and artificial data
illustrate the great flexibility that this framework offers.

For continuous data, Sklar's theorem ensures that the copula fully
describes the dependence structure separately from any marginal
properties; this allows the construction of a bivariate mixture model
in Example~\ref{motivatingExampleCont} that has Normal marginals and
can accommodate extreme tail dependence in the clusters. Such
flexibility allows the construction of mixture models that are capable
of producing a variety of exotic shapes (for example, star-shaped or
banana-shaped clusters), that are far from the cluster shapes that are
supported from contemporary proposals in the literature.

For discrete multivariate data, usual dependence measures are not
anymore margin-free, but the specification of the copula still
allows the easy construction of flexible multivariate mixture
models. This fact is used in Example~\ref{cognitive} where a
mixture of trivariate Binomial distributions with 4 alternative
copula specifications was used for the construction of performance
clusters for the students from their performance on a fraction
subtraction test.

Furthermore, using copulas one may define mixture models that
allow the joint modelling of mixed-domain variables. The
advantages of this approach are illustrated in Example~\ref{nba}
where a mixture model with components that have six Beta
distributions and one Gamma distribution as marginals each was
fitted to NBA data, allowing at the same time the use of a full
correlation specification amongst those.

\subsection{Model selection}
The ability to construct new copula families from known copulas
certainly adds great flexibility when constructing copula-based
mixture models for clustering. However, as mentioned in
Subsection~\ref{realvalued} it certainly does not simplify model
selection in a mixture model framework; if we limit ourselves to a
dictionary of $d$ copulas, then for finding the best model amongst
the models with $k$ components we need to fit and select the best
model from $\binom{d + k - 1}{k}$ models. Furthermore, if the
number of components is also considered as part of the model
selection exercise, then one needs to fit and compare $\sum_{k =
1}^K\binom{d + k - 1}{k}$ where $K$ is a preset maximum number of
components. Both these numbers increase quickly as either $K$ or
$d$ increase possibly making the model selection exercise
impractical.

In this respect, for data sets with real-valued observations,
copula-based mixture models were extended by introducing
component-wise parametric rotations and describing the ECM
algorithm that can fit these models. Example~\ref{rotexample}
illustrates that parametric rotations in two dimensions allow the
use of a single copula for capturing a range of dependence
structures, which would otherwise require the use of copulas with
different dependence properties.

The extension of the idea of rotations to many dimensions is
possible following exactly the same prescription as in two
dimensions, but using $p$-dimensional rotation matrices. An
accessible account of rotations in arbitrary dimensions can be
found in \citet{hanson:95}. Nevertheless, in order to incorporate
component-wise rotations in $p$ dimensions, $kp(p-1)/2$ extra
parameters are necessary (the number of components times the
number of free parameters in an $p$-dimensional orthogonal matrix)
which can become quickly impractical. Current work focuses on
using latent angular processes for the rotation angles which are
characterized by only a few parameters.



\subsection{Other special modelling settings and extensions}
Using the framework that is outlined in this paper one can construct
mixture models for the modelling of mixed-mode data; namely, data sets
that have some continuous, some discrete and some ordinal variables.
Despite the fact that this kind of data appears often, in practical
applications their joint treatment has been largely overlooked, mainly
because there are not any appropriate and easy to handle
models. Typically, models based on latent variables are considered for
such data \citep{Browne20122976} which may have limitations for
practical purposes because of assumptions like conditional
independence. A recent attempt for model-based clustering of mixed
mode data using the Gaussian copula can be found in the pre-print of
\citet{marbac:14}.



Furthermore, there are several practical scenarios, where the
marginal distributions need to be the same and the dependence
structure needs to be allowed to change. Such scenarios occur, for
example, in finance (regarding the behaviour of a portfolio when
information for the market comes), sports (scoring behaviour
depends on the current score), marketing (purchase frequency
patterns depend on household decomposition), etc. Different
dependence structures can be captured by different copulas and
hence mixtures of copulas with fixed marginals across components
can be used to cluster data with respect to their dependence
behavior.

Example~\ref{nba} and Example~\ref{cognitive} fitted mixture
models with components with exchangeable and unstructured
correlation matrices for the Gaussian copula. A wide-range of
parsimonious parameterizations between exchangeable and
unstructured correlation matrices can be obtained by adopting
ideas for parsimonious parameterizations in Normal mixture models
like the eigenvalue decomposition proposed in \citet{celeux:95}
and the factor analyzers proposed in \citet{mcnicholas:08}. These
can be directly applied to any copula family that is parameterized
in terms of a full variance-covariance matrix (like the Gausssian
and the $t$ copulas), allowing the comparison of a wide range of
parsimonious models. The study of such parsimonious
parameterizations and the implications on the cluster shapes for
various types of marginal distributions will be the focus of a
future study.

\subsection{Large dimensions}
More investigations are needed for the application of the
framework on scenarios with large $p$.  Simple copula families,
like Archimedean copulas, while attractive and easy to work with
in small dimensions, can have limited dependence structure (for
example, common dependence parameters, i.e., assuming that some
correlations are the same) in large dimensions.

As a starting point one may consider vine copulas \citep[see, for
example,][]{bedford:02} which use the fact that a $p$-dimensional
density can be decomposed into products of marginal densities and
bivariate copula-specified densities. This can lead to flexible
distributions with computationally tractable densities, at the expense
that the property of closure under marginalization is not satisfied in
general, and numerical integration is necessary for the calculation of
marginals (see Section~\ref{closure}). For discrete models one may use
the construction defined in \citet{panagiotelis:12} to construct
flexible multivariate discrete distributions.

\subsection{Computational effort}
The implementation of the fitting procedures described in
Section~\ref{modelfitting} has been done in R \citep{rcore:15}. The
code was written having in mind the ability to fit a diverse variety
of mixture models in terms of the choice of component copulas and
marginal combinations, instead of computational efficiency and
scalability. In this respect, the available implementation is nowhere
close to optimal regarding the latter. We also had to make convenience
choices and directly interface with other packages, including
\texttt{copula} \citep{copula:15} and \texttt{maxLik}
\citep{maxlik:15}. Such interfacing has introduced bottlenecks, due to
the necessary checks they need to be doing to the supplied inputs.

Keeping this in mind, Table~\ref{computetimes} lists the time (in
minutes) it took to fit the best models in
Example~\ref{motivatingExampleCont}, Example~\ref{nba},
Example~\ref{cognitive} and Example~\ref{rotexample}, the
characteristics of each model ($n$, $k$, $p$ and $q$), and the fitting
algorithm that has been used. All timings took place on an iMac (Late
2014) with a 4 GHz Intel Core i7 with Hyper-Threading enabled, and 32
GB of RAM memory, running R version 3.1.3. Parallelization across
components was used as described in Subsection~\ref{compute}. The
fitting time that is reported in Table~\ref{computetimes} is the
average from 10 identical repetitions of the fitting process. This is
done in an attempt to factor out as much as possible of the effect
that other OS-specific processes can have on timing. All computing
times shown can be drastically reduced by a slightly less stringent
termination criterion (see Subsection~\ref{EMsection}), and,
definitely, by a more optimised implementation of the fitting
procedures.

\begin{table}[t!]
  \begin{center}
    \begin{tabular}{lccccccc}
      \toprule
      Example & Algorithm & $n$ & $k$ & $p$ & $q$ & Iterations & Time \\
      \midrule
      Example~\ref{motivatingExampleCont} (artificial data) & ECM &
                                                                    800
                                & 4 & 2 & 23 & 37 & 1.68 \\
      Example~\ref{nba} (NBA) & ECM & 493 & 6 & 7 & 95 & 60  & 3.57 \\
      Example~\ref{rotexample} (artificial data, rotations)
              & ECM & 800 & 4 & 2 & 27 & 680 & 28.76 \\
      Example~\ref{cognitive} (cognitive diagnosis)  & EM & 536 & 6 &
                                                                      3
                                            & 29
                                            & 47 & 16.82 \\
      \bottomrule
    \end{tabular}
\end{center}
\caption{Computing times (in minutes) for fitting the best
  models in Example~\ref{motivatingExampleCont}, Example~\ref{nba},
  Example~\ref{cognitive} and Example~\ref{rotexample}.}
  \label{computetimes}
\end{table}

\subsection{Supplementary material}
Supplementary material extends Example~\ref{nba} to illustrate
that distinct sensible, transformations can lead to different
results. R scripts that reproduce the analyses undertaken in this
paper are available upon request to the authors.

\bibliographystyle{chicago}



\section*{Supplementary material}

# Model-based clustering using copulas with applications

Supplementary material

Ioannis Kosmidis and Dimitris Karlis

July 2, 2015

## 1 Transformations bounded- and mixed-domain variables

This section contains supplementary material for Section 4.2 ("Bounded- and mixed-domain variables") of the main text. The `mclust` (Fraley et al., 2012) R package has been used to find the best Normal mixture model according to BIC, for each of the following transformed versions of the data in Example 4.2

- `D-arcsine`: the data consists of the arcsine-transformed percentages ($\sin^{-1}(\sqrt{x})$), and the logarithm of the total points scored

- `D-logit`: the data consists of the logit-transformed percentages ($\log(x/(1-x))$), and the logarithm of the total points scored

- `D-probit`: the data consists of the probit-transformed percentages ($\Phi_1^{-1}(x)$), and the logarithm of the total points scored

For `D-arcsine` the best model was a 4-component Normal mixture model with VEV parameterization. On the other hand, for each of `D-logit` and `D-probit`, the best model was a 6-component Normal mixture model with VEV parameterization. Table 1 shows the adjusted Rand index for each of the possible pairs of clusterings corresponding to the aforementioned best models and the best model for the untransformed data set. The difference in the results across transformed data sets illustrates the point on the arbitrariness of choice of transformation that is made in Section 4.2 of the main text.

Furthermore, for each of the margin-transformed versions of the data, we conduct the same analysis as the one in Figure 4. The result is shown in Figure 1. A detailed explanation of how these scatterplots are constructed is provided in Example 4.2 of the main text. In terms of agreement to established metrics on characteristics of NBA players, the clustering from the selected copula-based mixture model for the original mixed-mode data set in Example 4.2 dominates all models for the transformed data.

| round | untransformed | D-arcsine | D-logit | D-probit |
|---|---|---|---|---|
| untransformed | 1.00 | 0.41 | 0.44 | 0.42 |
| D-arcsine | 0.41 | 1.00 | 0.48 | 0.48 |
| D-logit | 0.44 | 0.48 | 1.00 | 0.60 |
| D-probit | 0.42 | 0.48 | 0.60 | 1.00 |

Table 1: The adjusted Rand index for each of the possible pairs of clusterings corresponding to the best Normal mixture models (according to BIC) for the untransformed, `D-arcsine`, `D-logit` and `D-probit` data sets.

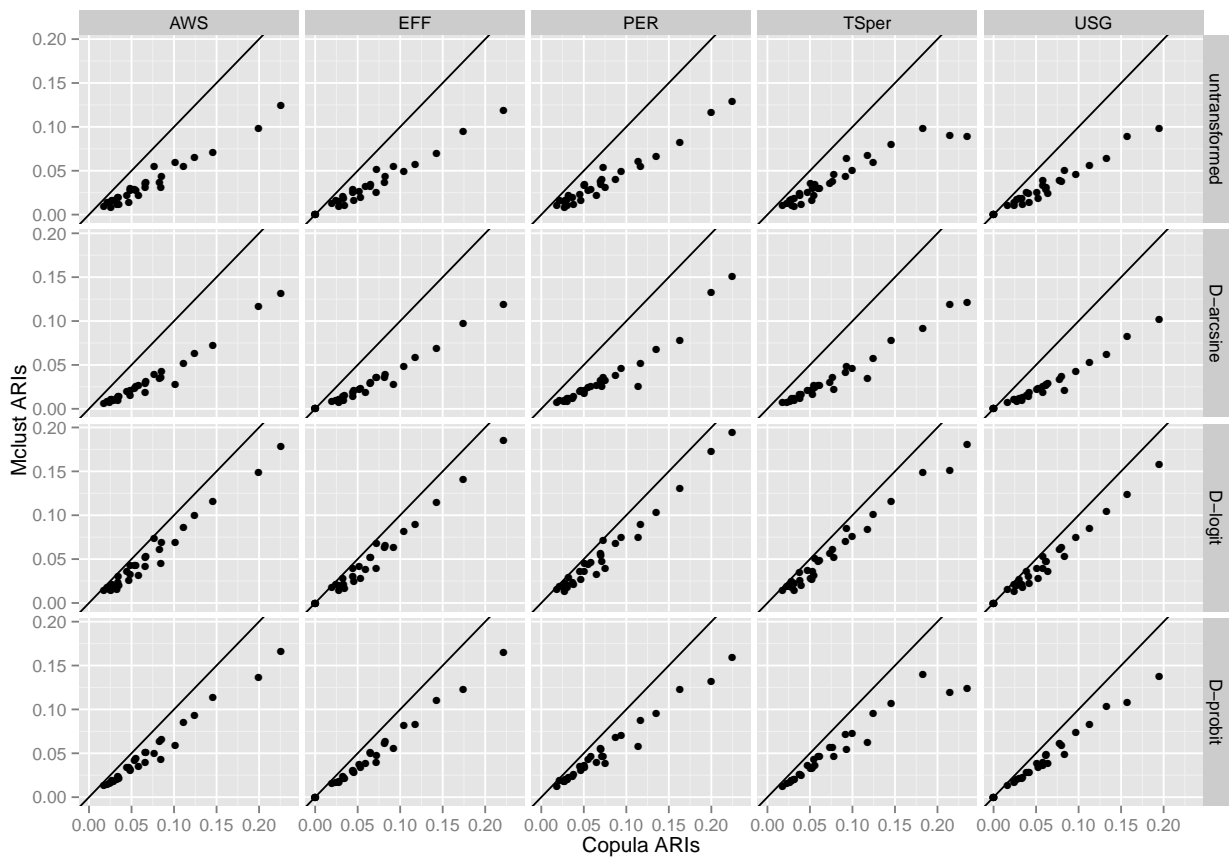

Figure 1: Clustering quality of best copula-based mixture model versus the best Normal mixture models for the data in Example 4.2 and their `D-arcsine`, `D-logit` and `D-probit` transformed versions. The dashed line is the $45^o$ line from the origin.




\end{document}